\documentclass[iop]{emulateapj}
\usepackage{amsfonts}
\usepackage{rotating}
\usepackage{multirow}

\shorttitle{The {\em NuSTAR\ } View of Nearby Compton-thick AGN}
\shortauthors{M. Balokovi\'{c} {\it et~al.}}

\def \etal {{\it et~al.\ $\!$}}
\def \nustar {{\em NuSTAR\ }}
\def \swift {{\em Swift\ }}
\def \swiftxrt {{\em Swift}/XRT\ }
\def \swiftbat {{\em Swift}/BAT\ }

\def \xmm {{\em XMM\ }}
\def \xmmnewton {{\em XMM-Newton\ }}
\def \chandra {{\em Chandra\ }}
\def \asca {{\em ASCA\ }}

\def \bepposax {{\em BeppoSAX\ }}
\def \integral {{\em INTEGRAL\ }}
\def \cgsflux {{erg~s$^{-1}$~cm$^{-2}$\,}}
\def \xspec {{\tt Xspec\ }}
\def \pexrav {{\tt pexrav\ }}

\def \mytorus {{\tt MYtorus\ }}
\def \msun {$M_{\odot}$}

\slugcomment{  }

\begin{document}

\title{The {\em NuSTAR\ } View of Nearby Compton-thick AGN:\\The Cases of NGC~424, NGC~1320 and IC~2560}

\author{
M.\,Balokovi\'{c}\altaffilmark{1}, 
A.\,Comastri\altaffilmark{2}, 
F.\,A.\,Harrison\altaffilmark{1},
D.\,M.\,Alexander\altaffilmark{3}, 
D.\,R.\,Ballantyne\altaffilmark{4},
F.\,E.\,Bauer\altaffilmark{5,6},
S.\,E.\,Boggs\altaffilmark{7}, 
W.\,N.\,Brandt\altaffilmark{8,9}, 
M.\,Brightman\altaffilmark{10}, 
F.\,E.\,Christensen\altaffilmark{11}, 
W.\,W.\,Craig\altaffilmark{7,12}, 
A.\,Del\,Moro\altaffilmark{3},  
P.\,Gandhi\altaffilmark{3},
C.\,J.\,Hailey\altaffilmark{13},  
M.\,Koss\altaffilmark{14,15}, 
G.\,B.\,Lansbury\altaffilmark{3}, 
B.\,Luo\altaffilmark{8,9},
G.\,M.\,Madejski\altaffilmark{16}, 
A.\,Marinucci\altaffilmark{17},
G.\,Matt\altaffilmark{17},
C.\,B.\,Markwardt\altaffilmark{18},
S.\,Puccetti\altaffilmark{19,20}
C.\,S.\,Reynolds\altaffilmark{21,22},   
G.\,Risaliti\altaffilmark{23,24}, 
E.\,Rivers\altaffilmark{1}, 
D.\,Stern\altaffilmark{25}, 
D.\,J.\,Walton\altaffilmark{1}, 
W.\,W.\,Zhang\altaffilmark{18}
}

\altaffiltext{1}{Cahill Center for Astronomy and Astrophysics, Caltech, Pasadena, CA 91125, USA}
\altaffiltext{2}{INAF Osservatorio Astronomico di Bologna, via Ranzani 1, I-40127, Bologna, Italy}
\altaffiltext{3}{Department of Physics, Durham University, Durham DH1 3LE, UK}
\altaffiltext{4}{Center for Relativistic Astrophysics, School of Physics, Georgia Institute of Technology, Atlanta, GA 30332, USA}
\altaffiltext{5}{Instituto de Astrof\'{i}sica, Facultad de F\'{i}sica, Pontificia Universidad Cat\'{o}lica de Chile 306, Santiago 22, Chile} 
\altaffiltext{6}{Space Science Institute, 4750 Walnut Street, Suite 205, Boulder, CO 80301, USA}
\altaffiltext{7}{Space Sciences Laboratory, University of California, Berkeley, CA 94720, USA}
\altaffiltext{8}{Department of Astronomy and Astrophysics, The Pennsylvania State University, 525 Davey Lab, University Park, PA 16802, USA}
\altaffiltext{9}{Institute for Gravitation and the Cosmos, The Pennsylvania State University, University Park, PA 16802, USA}
\altaffiltext{10}{Max-Planck-Institut f{\" u}r extraterrestrische Physik, Giessenbachstrasse 1, D-85748 Garching bei M{\" u}nchen, Germany}
\altaffiltext{11}{DTU Space, National Space Institute, Technical University of Denmark, Elektrovej 327, DK-2800 Lyngby, Denmark}
\altaffiltext{12}{Lawrence Livermore National Laboratory, Livermore, CA 94550, USA}
\altaffiltext{13}{Columbia Astrophysics Laboratory, Columbia University, New York, New York 10027, USA}
\altaffiltext{14}{Institute for Astronomy, University of Hawaii, 2680 Woodlawn Drive, Honolulu, HI 96822, USA}
\altaffiltext{15}{Institute for Astronomy, Department of Physics, ETH Zurich, Wolfgang-Pauli-Strasse 27, CH-8093 Zurich, Switzerland}
\altaffiltext{16}{Kavli Institute for Particle Astrophysics and Cosmology, SLAC National Accelerator Laboratory, Menlo Park, CA 94025, USA}
\altaffiltext{17}{Dipartimento di Matematica e Fisica, Universit{\` a} degli Studi Roma Tre, via della Vasca Navale 84, I-00146 Roma, Italy}
\altaffiltext{18}{NASA Goddard Space Flight Center, Greenbelt, MD 20771, USA}
\altaffiltext{19}{ASI-Science Data Center, via Galileo Galilei, I-00044 Frascati, Italy}
\altaffiltext{20}{INAF Osservatorio Astronomico di Roma, via Frascati 33, I-00040 Monteporzio Catone, Italy}
\altaffiltext{21}{Department of Astronomy, University of Maryland, College Park, MD, USA}
\altaffiltext{22}{Joint Space-Science Institute, University of Maryland, College Park, MD, USA}
\altaffiltext{23}{INAF Osservatorio Astrofisico di Arcetri, Largo E. Fermi 5, I-50125 Firenze, Italy}
\altaffiltext{24}{Harvard-Smithsonian Center for Astrophysics, 60 Garden St., Cambridge, MA 02138, USA}
\altaffiltext{25}{Jet Propulsion Laboratory, California Institute of Technology, Pasadena, CA 91109, USA}

\begin{abstract}
We present X-ray spectral analyses for three Seyfert~2 active galactic nuclei, NGC~424, NGC~1320, and IC~2560, observed by \nustar in the 3--79~keV band. The high quality hard X-ray spectra allow detailed modeling of the Compton reflection component for the first time in these sources. Using quasi-simultaneous \nustar and \swiftxrt data, as well as archival \xmmnewton data, we find that all three nuclei are obscured by Compton-thick material with column densities in excess of $\sim5\times10^{24}$~cm$^{-2}$, and that their X-ray spectra above 3~keV are dominated by reflection of the intrinsic continuum on Compton-thick material. Due to the very high obscuration, absorbed intrinsic continuum components are not formally required by the data in any of the sources. We constrain the intrinsic photon indices and the column density of the reflecting medium through the shape of the reflection spectra. Using archival multi-wavelength data we recover the intrinsic X-ray luminosities consistent with the broadband spectral energy distributions. Our results are consistent with the reflecting medium being an edge-on clumpy torus with a relatively large global covering factor and overall reflection efficiency of the order of 1\%. Given the unambiguous confirmation of the Compton-thick nature of the sources, we investigate whether similar sources are likely to be missed by commonly used selection criteria for Compton-thick AGN, and explore the possibility of finding their high-redshift counterparts.
\end{abstract}


\section{Introduction} 

\label{sec:intro}

It is well established that a significant fraction of active galactic nuclei (AGN) are intrinsically obscured by gas and dust surrounding the central supermassive black holes (SMBH). Obscured AGN are needed to explain the $\sim$30~keV peak of the Cosmic X-ray Background (CXB; e.g.,\,\citealt{churazov+2007,frontera+2007,ajello+2008,moretti+2009}), however, their space density is observationally poorly constrained. AGN obscured by gas with column density of $N_{\rm H}\lesssim1.5\times10^{24}$~cm$^{-2}$ have been identified in large numbers in deep soft X-ray ($<\!10$~keV) surveys \citep{brandt+hasinger-2005}, which are powerful means for identifying the bulk of the AGN population at high redshift and thus providing valuable constraints on the growth history of SMBH (e.g.\,\citealt{lafranca+2005,aird+2010}). However, the heavily obscured, Compton-thick sources ($N_{\rm H}>1.5\times10^{24}$~cm$^{-2}$; see, e.g.\,\citealt{comastri-2004} for a review) required by the CXB models remain elusive.

Recent surveys with the hard X-ray ($>\!10$~keV) telescopes \swiftbat ({\em Burst Alert Telescope}; \citealt{gehrels+2004}) and \integral ({\em International Gamma-Ray Astrophysics Laboratory}; \citealt{winkler+2003}) indicate that in the local Universe the fraction of obscured AGN (with $N_{\rm H}>10^{22}$~cm$^{-2}$) is approximately 80\%, while Compton-thick sources likely contribute about 20\% of the total number of AGN (estimated from the observed $\lesssim$10\% fraction corrected for survey completeness, e.g., \citealt{malizia+2009,burlon+2011}). Obscured AGN therefore contribute significantly to the local supermassive black hole space density \citep{marconi+2004} and may be even more important at earlier epochs \citep{lafranca+2005,ballantyne+2006,treister+urry-2006,brightman+ueda-2012,iwasawa+2012}. The peak of the CXB at $\sim$30~keV can be reproduced by invoking a significant number of Compton-thick sources at moderate redshift (e.g.,\,\citealt{gilli+2007,treister+2009,ballantyne+2011,akylas+2012}), however, to date only a few percent of the CXB has actually been resolved at its peak energy (e.g.,\,\citealt{ajello+2008,bottacini+2012}). The distribution of the obscuring column density and the degeneracy in relative contributions of absorption- and reflection-dominated hard X-ray spectra are therefore poorly constrained with the current data.

A primary goal of the \nustar ({\em Nuclear Spectroscopic Telescope Array}) hard X-ray mission is to study the evolution of obscuration in AGN at $0\!<\!z\!<\!2$ through its comprehensive extragalactic survey program \citep{harrison+2013}. In addition to blank-field observations (Mullaney \etal, {\it in prep.}; Civano \etal, {\it in prep.}), the program includes a survey of known sources selected from the \swiftbat catalog with two goals: (i) obtain high-quality spectroscopy of the nearby \swiftbat$\!\!$-selected AGN, and (ii) perform a wide-field search for serendipitous background sources \citep{alexander+2013}. In this paper we present observations and modeling of the hard X-ray spectra of three local AGN: NGC~424, NGC~1320, and IC~2560, two of which are selected from the program outlined above. All three show spectra dominated by reflection from cold, distant, Compton-thick material, the properties of which are impossible to fully constrain using only soft X-ray data. These Compton-thick AGN demonstrate how \nustar spectroscopy of the nearby targets can characterize their X-ray properties better than previously possible, and how the new constraints may lead to improved understanding of both local AGN and their distant counterparts. Ultimately, the \nustar surveys will allow us to directly determine the fraction of Compton-thick sources in the AGN population and the distribution of the obscuring column density for heavily obscured AGN.

This paper is organized as follows. In \S\,2 we present the target selection, and the new data obtained from quasi-simultaneous observations with \nustar and {\em Swift}. \S\,3.1 and \S\,3.2 demonstrate that reflection is the dominant component of the hard X-ray spectra of these three AGN. \S\,3.3 provides a more detailed spectral analysis including the \xmm data. A comparison with the previously published X-ray results, as well as a discussion of the multi-wavelength properties and constraints on the AGN geometry, is presented in \S\,4. We summarize our results in \S\,5. In this work we use standard cosmological parameters ($h_0=0.7$, $\Omega_{\Lambda}=0.73$) to calculate distances. Unless noted otherwise, all uncertainties are given as 90\% confidence intervals.

\section{Target Selection and Observations} 

\label{sec:obs+data}

\subsection{Target Selection} 

The \nustar Extragalactic Survey program includes a wide-field shallow component (average exposure of 20~ks) in which the observatory is pointed towards a known AGN previously detected with {\swiftbat$\!\!$,} or selected because of high obscuration inferred from soft X-ray ($<10$~keV) observations. The wide field of view of the \swiftbat instrument and its nearly uniform coverage of the whole sky down to a sensitivity of $\gtrsim1.0\times10^{-11}$~erg~s$^{-1}$~cm$^{-2}$ in the 14--195~keV band \citep{baumgartner+2013-swiftbat70}, provide a reasonable sample of predominantly local ($z\sim0.03$) AGN. Its more uniform and deeper exposure of the sky away from the Galactic plane compared to \integral makes the \swiftbat survey source catalog an excellent starting point for selecting targets for more detailed spectroscopic studies, such as possible with {\nustar$\!\!$.} The targets were selected for \nustar observations from the catalog utilizing 54 months of BAT operation \citep{cusumano+2010-swiftbat54}. Unless prevented by technical constraints, all \nustar targets in this program receive on average 7~ks of quasi-simultaneous coverage (with delay of $\lesssim$1 day) in the soft X-ray band from the \swiftxrt$\!\!$ in order to enable spectral analysis over the broad 0.3--79~keV band.

Two of the targets presented here, NGC~424 (Tololo~0109--383) and NGC~1320, were selected from the 54-month \swiftbat catalog.\footnote{Due to different methodology and low significance, NGC~1320 is not included in the latest 70-month catalog \citep{baumgartner+2013-swiftbat70}} The third target, IC~2560, was selected from a sample of relatively faint AGN with some indication of Compton-thick obscuration from previous observations (e.g.\,\citealt{risaliti+1999,tilak+2008}). Soft X-ray spectroscopy, as well as multi-wavelength data, indirectly suggest that NGC~424 and NGC~1320 are also likely to be Compton-thick \citep{collinge+brandt-2000,marinucci+2011,brightman+nandra-2011a}. The selection of the sample presented here is based on a basic spectral analysis of all \nustar 20-ks snapshot observations of AGN up to May 2013. Out of 34 observed AGN we selected 3 that show the most prominent Compton reflection component signature: very hard spectrum ($\Gamma<1$, assuming a simple power law model), strong Compton hump (high reflection fraction, $R>10$, assuming the simplest reflection model) and iron emission (large equivalent width of the neutral iron K$\alpha$ line, $\gtrsim1$~keV). This is not a uniformly selected, statistically complete sample -- we will address such samples in future work. However, the hard X-ray properties of these three targets can be considered representative of a larger class of heavily obscured AGN, which make up approximately 10\% of the sample of nearby AGN being surveyed with \nustar$\!\!$. The observed spectra of all three sources are shown in Figure~\ref{fig:spectra+ratios}. Some basic data on the targets is summarized in Table~\ref{tab:basic_data}.

\begin{deluxetable}{r c c c} 

\tabletypesize{\scriptsize}
\tablecaption{ Basic data on the AGN presented here. \label{tab:basic_data} }

\tablehead{
  \colhead{ } &
  \colhead{\multirow{2}{*}{\bf NGC~424}} &
  \colhead{\multirow{2}{*}{\bf NGC~1320}} &
  \colhead{\multirow{2}{*}{\bf IC~2560}} \\
  \colhead{ } &
  \colhead{ } &
  \colhead{ } &
  \colhead{ } }

\startdata

Galaxy Type\,\tablenotemark{a} 					& SB0/a 				& Sa 					& SBb \\
AGN Type\,\tablenotemark{a}						& Sy\,1/Sy\,2		 	& Sy\,2 				& Sy\,2 \\
$M_{\rm BH}$ [ \msun ]\,\tablenotemark{b}		& $6.0\times10^7$		& $1.5\times10^7$		& $2.9\times10^6$ \\
Redshift ($z$)\,\tablenotemark{c}				& 0.0117				& 0.0091				& 0.0096 \\
$d_L$ [ Mpc ]\,\tablenotemark{c}					& 50.6					& 39.1					& 41.4 \\
$N_{\rm H,G}$ [ cm$^{-2}$ ]\,\tablenotemark{d}	& $1.7\times10^{20}$	& $4.3\times10^{20}$ 	& $6.8\times10^{20}$ \\

\enddata

\tablenotetext{a}{Summary of classifications from the NASA Extragalactic Database (NED; \texttt{http://ned.ipac.caltech.edu/}).}
\tablenotetext{b}{Black hole mass from \citet{greenhill+2008} and \citet{bian+gu-2007}.}
\tablenotetext{c}{Adopted redshift and luminosity distance (calculated assuming $h_0=0.7$, $\Omega_{\Lambda}=0.73$) based on published measurements available through NED. Note that the distances used in the literature differ $\lesssim10$\% for NGC~424 and NGC~1320, and up to 40\% for IC~2560.}
\tablenotetext{d}{Galactic column density averaged between \citet{dickey+lockman-1990} and \citet{kalberla+2005}. \vspace{0.5cm}}

\end{deluxetable} 

\subsection{NuSTAR Data} 

\label{sec:obs+data-nustar}

NGC~1320 was observed on two occasions: an initial 15-ks snapshot and additional follow-up to improve the signal-to-noise ratio for detailed spectral anaysis. As a first step in our analysis we check for variability between the observations and find that they are consistent with no change; hereafter we analyse them jointly, but without co-adding. The other two sources were observed once each. Table~\ref{tab:obslog-xray} gives a summary of all \nustar observations. The raw data were reduced using the NuSTARDAS software package (version~1.2.1), distributed with the HEAsoft package by the NASA High Energy Astrophysics Archive Research Center (HEASARC). The raw events were cleaned and filtered for South Atlantic Anomaly (SAA) passages using the \texttt{nupipeline} task. The cleaned events were further processed for each of the two focal plane modules (FPMA and FPMB) using the \texttt{nuproducts} task, which generates the spectra and the corresponding response files. These procedures are presented in detail in \citet{perri+2013}.

Spectra for all of the sources were extracted from circular regions 40\arcsec\ in radius, centered on the peaks of the point-source images. The background spectra were extracted from regions encompassing the same detector as the source,\footnote{In each module the focal plane consists of 4 detectors; for details, see \citet{harrison+2013}.} excluding the circular region 50\arcsec\ around the source. The background region sampling the same detector as the source provides the best estimate of the underlying background. For IC~2560 the background was extracted from two adjacent detectors due to its position in the focal plane. All fluxes reported in this paper have been automatically corrected for the finite extraction aperture using the best point spread function model currently available. We do not use \nustar data below 3~keV, since the calibration is currently uncertain in that energy range. The upper end of the bandpass is mostly limited by photon statistics and the \nustar instrumental background. All \nustar spectra are binned to a minimum of 20~photons per bin using HEAsoft task \texttt{grppha}.

\subsection{Swift/{\em XRT} Data} 

Each \nustar observation was accompanied by a short observation with \swift$\!\!$, typically delayed by less than 24~hours. The purpose of these observations was to provide coverage on the soft X-ray end of the spectrum, where the \nustar sensitivity drops off, and to facilitate a comparison of the soft X-ray flux with the data available in the literature. Since the sources are not expected to be highly variable on timescales of hours, quasi-simultaneous exposures with \nustar and \swiftxrt provide a broadband snapshot covering the range from approximately 0.5 to 70~keV. The \swiftxrt observations were performed in the Photon Counting mode \citep{hill+2004,burrows+2005}. The data were reduced using the task \texttt{xrtpipeline} (version~0.12.6), which is a part of the XRT Data Analysis Software (XRTDAS) within HEAsoft. Spectra were extracted from circular regions 20\arcsec\ in radius centered on the targets and the backgrounds were extracted from large annular source-free regions around them. We used response file \texttt{swxpc0to12s6\_20010101v013.rmf} from the \swiftxrt calibration database, while auxilliary response files were generated using the task \texttt{xrtmkarf}. Table~\ref{tab:obslog-xray} provides the complete list of observations. Unfortunately, the observation of IC~2560 was too short to yield a detection in the \swiftxrt band, so we use it here only to place an upper limit on the soft X-ray emission. Due to low count statistics, the \swiftxrt spectra are binned to a minimum of 10~photons per bin.

\subsection{Archival XMM-Newton Data} 

\label{sec:archival_data}

In addition to the quasi-simultaneous \nustar and \swiftxrt data, we use archival data from \xmmnewton to additionally verify our models. The \xmm spectra are the highest-quality soft X-ray spectra currently available for these sources. Descriptions of the data and details regarding their reduction can be found in \citet{marinucci+2011}, \citet{brightman+nandra-2011a} and \citet{tilak+2008} for NGC~424, NGC~1320 and IC~2560, respectively.

\begin{figure} 
\begin{center} 
\includegraphics[width=0.97\columnwidth]{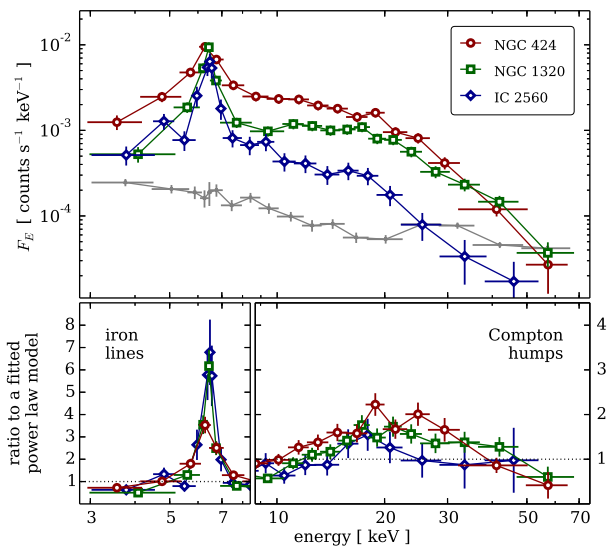}
\caption{ {\it Upper panel:} The observed \nustar spectra of NGC~424 (red circles), NGC~1320 (green rectangles) and IC~2560 (blue diamonds). Spectra for two focal plane modules have been co-added, and only the longer observation of NGC~1320 is shown here for clarity. The typical background level and its uncertainty are shown by the filled grey symbols and lines. {\it Lower panels:} Ratio of the spectra and a simple power-law model fitted to each spectrum (symbols as in the panel above). Note that the right panel is rescaled vertically with respect to the left panel by a factor of 2 in order to better show the Compton humps. \label{fig:spectra+ratios}}
\end{center} 
\end{figure} 

\begin{deluxetable*}{cccccc} 

\tabletypesize{\scriptsize}
\tablecaption{Summary of the quasi-simultaneous \nustar and \swift observations.
\label{tab:obslog-xray} }

\tablehead{
  \colhead{\multirow{2}{*}{Target}} &
  \colhead{Sequence} &
  \colhead{Start Time} &
  \colhead{Duration} &
  \colhead{Exposure} &
  \colhead{Count Rate\tablenotemark{a}} \\
  \colhead{} &
  \colhead{ID} &
  \colhead{[ UTC ]} &
  \colhead{[ ks ]} &
  \colhead{[ ks ]} &
  \colhead{[ $10^{-2}$ counts s$^{-1}$ ]} }

\startdata

\cutinhead{\nustar Observations}
NGC 1320 & 60061036002 & 2012-Oct-25 21:50 & 25.7 & 14.5 & 2.8$\pm$0.2 / 2.4$\pm$0.2 \\
NGC 424 & 60061007002 & 2013-Jan-26 06:35 & 26.7 & 15.5 & 4.7$\pm$0.2 / 4.6$\pm$0.2 \\
IC 2560 & 50001039002 & 2013-Jan-28 22:05 & 43.5 & 23.4 & 1.1$\pm$0.1 / 1.1$\pm$0.1 \\
NGC 1320 & 60061036004 & 2013-Feb-10 07:15 & 49.9 & 28.0 & 3.0$\pm$0.1 / 2.6$\pm$0.1 \\
\cutinhead{\swiftxrt Observations}
NGC 1320 & 00080314001 & 2012-Oct-26 02:48 & 75.6 & 6.8 & 1.04$\pm$0.03 \\
NGC 424 & 00080014001 & 2013-Jan-26 06:34 & 23.9 & 6.6 & 2.55$\pm$0.03 \\
IC 2560 & 00080034001 & 2013 Jan 29 17:50 & 2.0 & 2.0 & $<$0.42 \\
NGC 1320 & 00080314002 & 2013-Feb-10 07:19 & 23.9 & 6.6 & 1.04$\pm$0.03 \\

\enddata

\tablenotetext{a}{Count rates for \nustar modules FPMA and FPMB (3--79~keV), or \swiftxrt (0.3--10~keV). \vspace{0.5cm}}

\end{deluxetable*} 

\begin{figure*} 
\begin{center} 
\includegraphics[width=1.9\columnwidth]{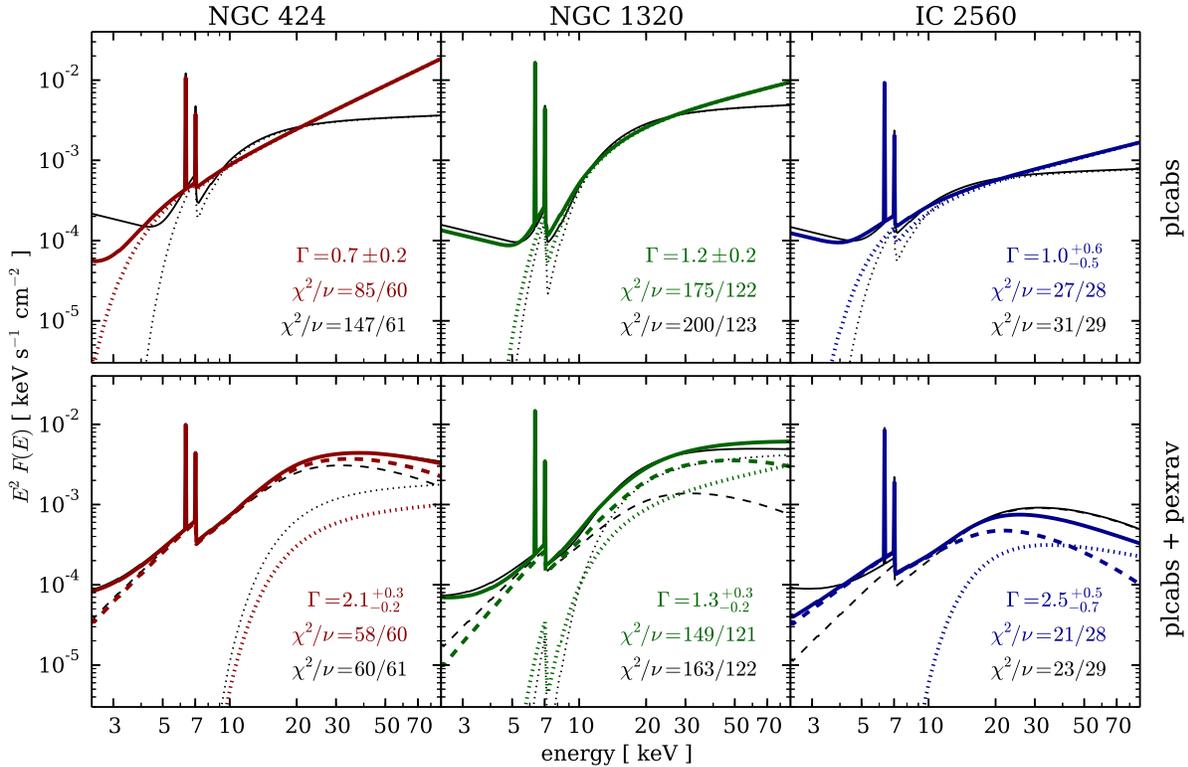}
\caption{ Simple approximate models fitted to the NGC~424, NGC~1320 and IC~2560 \swiftxrt and \nustar data (see \S\,\ref{sec:modeling1} for details). Thick colored lines show models for which the photon index was determined from the data; the best-fit value, its 90\% confidence interval and the best fit $\chi^2$ are given in the lower right corner of each panel. The thin black lines show the same models with the assumption of $\Gamma=1.9$; its $\chi^2$ is given in black letters. With the solid lines we show the total model, while the dashed and dotted lines show the reflection and the transmission components, respectively. {\it Upper panels:} Transmission-only models based on \texttt{plcabs}. {\it Lower panels:} Two-component models based on the \texttt{plcabs} and \texttt{pexrav} components. }
\label{fig:ss1}
\end{center} 
\end{figure*} 

\section{Modeling of the X-ray Spectra} 

The \nustar hard X-ray spectra (3--70~keV) of NGC~424, NGC~1320 and IC~2560 are qualitatively similar, as demonstrated in Figure~\ref{fig:spectra+ratios}. In the lower panel the figure we also show the ratios of the spectra to their respective best-fit power law model simply to highlight the most important features. The best-fit photon indices in all three cases are lower than unity; these fits are rather poor (reduced $\chi^2\!\gtrsim$3) and intended only for demonstration. The spectra exhibit very hard continua with a convex shape broadly peaking around 20~keV, and a prominent emission feature at 6.4~keV, matching the rest-frame energy of the neutral iron K$\alpha$ emission line. The hard effective photon indices and the structure of the residuals reveal the presence of a strong X-ray reflection component in the \nustar spectra. The prominent neutral iron line arising from fluorescence and the broadly peaked Compton hump in the 20--30~keV region are typical signatures of such a component (e.g.,\,\citealt{ghisellini+1994,matt+2000,matt+2003b}).

Detailed models of the soft X-ray spectra, which is composed of a combination of Thomson-scattered AGN light, plasma ionized by the AGN and star formation, are not the focus of this paper. We refer the reader to \citet{marinucci+2012}, \citet{brightman+nandra-2011b} and \citet{tilak+2008} for more details on such models. In the analysis presented here, a simple power law is used to approximate the contribution of the soft component(s) to the spectra above 3~keV. A good phenomenological model for the \swiftxrt soft X-ray data (0.3--3~keV) for both NGC~424 and NGC~1320 is a power law with $\Gamma_s\approx2.7$ ($\Gamma_s=2.5\pm0.8$ and $\Gamma_s=2.7\pm0.4$, respectively). We adopt this average value and keep it fixed in all models, varying only the normalization, in order to avoid the degeneracy associated with its large uncertainty. For all three targets we verify that this value is consistent with the higher-quality \xmmnewton data. The \xmm data clearly require a more complex spectral model in order to fit the data well, as additional fine structure on top of the slope is apparent in the $<3$~keV residuals. However, the simple power law represents a good approximation. More detailed modeling is not warranted for the hard X-ray analysis presented in this paper.

In the following subsections we present results from applying three different types of hard X-ray spectral models. First we apply simple, phenomenological models, which have been extensively used in previous work (\S\,\ref{sec:modeling1}). We also apply physically motivated torus models (\S\,\ref{sec:modeling2}) and reflection-only models (\S\,\ref{sec:modeling3}). We use \xspec version 12.8.1 \citep{arnaud+1996} for all our modeling. In addition, we: take into account redshifts and Galactic absorption column density listed in Table~\ref{tab:basic_data}; assume a contribution from a soft power law component ($\Gamma_s=2.6$) with free normalization; assume Solar abundances and a high-energy cutoff in the nuclear continuum at 200~keV;\footnote{We show later in \S\,\ref{sec:discussion-previous} that the NGC~424 data are consistent with this value; for the other two targets this parameter is unconstrained. This choice is consistent with the recent literature, e.g.,\,\citet{ballantyne-2014} and \citet{malizia+2014}.} leave the cross-normalization constants between instruments to vary freely. We perform all parameter optimizations using the Cash statistic \citep{cash-1979}, but report the $\chi^2$ values of each best fit due to their straightforward interpretability.

\begin{figure*} 
\begin{center} 
\includegraphics[width=1.9\columnwidth]{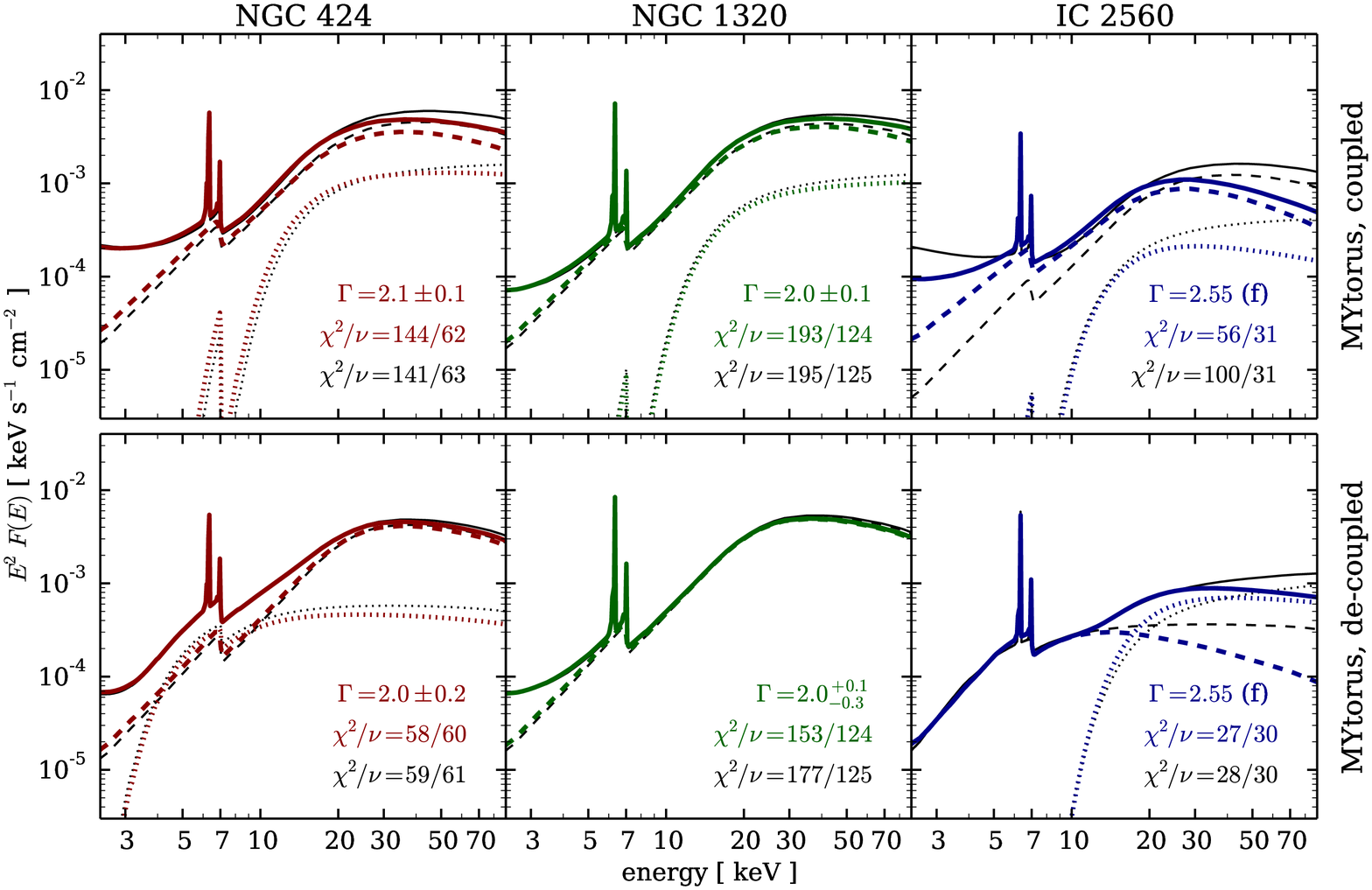}
\caption{ Torus models fitted to the NGC~424, NGC~1320 and IC~2560 \swiftxrt and \nustar data (see \S\,\ref{sec:modeling2} for details). Thick colored lines show models for which the photon index was determined from the data; the best-fit value, its 90\% confidence interval and the best fit $\chi^2$ are given in the lower right corner of each panel. The thin black lines show the same models with the assumption of $\Gamma=1.9$; its $\chi^2$ is given in black letters. With the solid lines we show the total model, while the dashed and dotted lines show the reflection and the transmission components, respectively. {\it Upper panels:} Literal torus models represented by \mytorus in the coupled mode, with equatorial column density fixed to $5\times10^{24}$~cm$^{-2}$. {\it Lower panels:} Generalized two-component models based on de-coupled components of \mytorus$\!\!$. }
\label{fig:ss2}
\end{center} 
\end{figure*} 

\subsection{Phenomenological Models} 

\label{sec:modeling1}

We first fit an absorbed power law model using the \xspec component \texttt{plcabs} \citep{yaqoob-1997}. This model represents an absorbed power-law spectrum including the effects of Compton scattering. This model is approximated in order to be computationally fast, and it is also limited to $N_{\rm H}<5\times10^{24}$~cm$^{-2}$. In all three cases the fits are poor and show a strong narrow residual feature around 6.4~keV. The addition of two unresolved Gaussian components ($\sigma=10^{-3}$~keV) at 6.40 and 7.06~keV, corresponding to neutral iron K$\alpha$ and K$\beta$ lines, improves the fits significantly. Although the reduced $\chi^2$ ($\chi^2/\nu$, where $\nu$ is the number of degrees of freedom) reaches $\simeq$1 for the case of IC~2560 and $\simeq$1.5 for the other two AGN, the fits are difficult to justify physically. The best fits are all qualitatively the same: the intrinsic power law with a photon index which tends to $\Gamma\lesssim1$ is absorbed by a column density of $10^{23-24}$~cm$^{-2}$, and the equivalent width of the Gaussian component at the energy of the iron K$\alpha$ line exceeds 2~keV. These values of $\Gamma$ are much harder than typical for the coronal continuum of AGN, while the equivalent width of the iron line strongly indicates presence of a reflection component. The photon indices can be assumed to take on a typical value of 1.9 at a cost of increasing the $\chi^2$, however the best fits remain qualitatively the same. They are shown for each of the three AGN in the top panels of Figure~\ref{fig:ss1}.

We next add a reflection component, which we approximate using the {\tt pexrav} model \citep{magdziarz+zdziarski-1995}. This model produces the reflected continuum of an infinite slab of infinite optical depth, and is therefore only an approximation for the reflection off a distant torus. We apply this component to produce only the reflected continuum. The incident spectrum is set to be the same intrinsic cut-off power law as in the {\tt plcabs} component. We start our fitting procedure with both the transmitted and the reflected component. We also include narrow Gaussian components at 6.40 and 7.06~keV and the soft $\Gamma=2.7$ power law. The basic result of the fitting is that the contribution of the transmitted components are minor for all three AGN: the data requires either very high absorption column ($N_{\rm H,A}\gtrsim5\times10^{24}$~cm$^{-2}$, which is the upper limit of the \texttt{plcabs} model), or zero normalization.

For NGC~424 the best-fit column density is $N_{\rm H,A}\gtrsim5\times10^{24}$~cm$^{-2}$ and the intrinsic power law continuum slope is $\Gamma=2.1_{-0.2}^{+0.3}$ ($\chi^2/\nu=58/60$). If we fix the absorption column density at a lower value, the normalization of the transmitted component decreases until it becomes consistent with zero for $N_{\rm H,A}=2\times10^{24}$~cm$^{-2}$. If the transmitted component is removed from the model altogether, the best fit ($\chi^2/\nu=59/61$) is found for $\Gamma=1.71\pm0.09$. We find qualitatively similar results for IC~2560. For a fixed $N_{\rm H,A}=5\times10^{24}$~cm$^{-2}$ the best-fit photon index is $2.5_{-0.7}^{+0.5}$ ($\chi^2/\nu=21/28$). For either a lower $N_{\rm H,A}$ or a lower $\Gamma$, the {\tt plcabs} component vanishes, and the best fit is found for a \pexrav$\!\!$-only model ($\chi^2/\nu=22/29$) with $\Gamma=2.2_{-0.4}^{+0.3}$. In the case of NGC~1320 the best-fit model ($\chi^2/\nu=149/121$) is dominated by the reflection component, but does include a transmitted power-law component with $\Gamma=1.3_{-0.2}^{+0.3}$ and $N_{\rm H,A}=(2_{-1}^{+2})\times10^{24}$~cm$^{-2}$. Simply removing the latter component degrades the fit only to $\chi^2/\nu=152/123$. An alternative model, dominated by the \texttt{plcabs} component above 10~keV, can be found with the assumption of $\Gamma=1.9$ ($\chi^2/\nu=163/122$).

The models presented in this subsection are summarized in Figure~\ref{fig:ss1}. For all three AGN we find that a statistically good description of their hard X-ray spectra can be achieved using models consisting almost entirely of reflection components. However, with the quality of the data acquired in short 20-ks exposures it is not possible to exclude a minor contribution from a transmitted intrinsic continuum. The large equivalent width of the iron lines point towards strong reflection and essentially rule out the possibility that the hard X-ray spectrum is primarily due to the transmission of the intrinsic continuum through mildly Compton-thick material. We further examine a set of more appropriate physically motivated models.

\subsection{Torus Models} 

\label{sec:modeling2}

An improvement over the \pexrav approximation, which assumes infinite optical depth, is offered by theoretical models that use Monte Carlo simulations of the propagation of X-ray photons through material of finite optical depth in a physically motivated geometry. The first \xspec model of that kind is \mytorus$\!\!\!$~\footnote{We use the version of \mytorus model that is publicly available at \texttt{http://www.mytorus.com}. Specifically, we use the tables calculated with a primary power law with a cutoff at 200~keV.} \citep{murphy+yaqoob-2009,yaqoob-2012}. The basis of the \mytorus model is a literal torus with a 60$^{\circ}$ half-opening angle. It consists of two main spectral components: a transmitted continuum component (formally called zeroth-order continuum, {\tt MYTZ}, by the authors of the model) and a scattered one ({\tt MYTS}; also referred to as reprocessed, or reflected). The former is produced by scattering photons {\em away} from the line of sight, while the latter is formed by photons scattered {\em into} the line of sight of the distant observer.

We start with the complete \mytorus model, which is characterized by a single column density ($N_{\rm H,R}=N_{\rm H,A}=N_{\rm H}$, corresponding to the column density in the equatorial plane of the torus). The internal normalizations between the components  are fixed. The first model we test is an edge-on torus with inclination fixed at 90$^{\circ}$. This model does not fit any of the \nustar and \swift data considered here: the reduced $\chi^2$ values do not get any lower than 2--3. Note that these models are transmission-dominated and therefore formally similar (but physically more appropriate) to the {\tt plcabs}-only model examined in \S\,\ref{sec:modeling1}.

Next we fit for the inclination angle of the torus under different assumptions of the equatorial column density, $N_{\rm H}$, since a straightforward fit for both of those parameters is highly degenerate. The results are again qualitatively the same for all three AGN, regardless of $N_{\rm H}$: the best-fit inclination angles are found to be close to $60^{\circ}$, matching the opening angle of the torus. For example, at $N_{\rm H}=5\times10^{24}$~cm$^{-2}$ (shown in the upper panels of Figure~\ref{fig:ss2}), the best-fit inclinations are $69_{-4}^{+5}$, $68_{-2}^{+3}$ and $66_{-4}^{+7}$ degrees for NGC~424, NGC~1320 and IC~2560, respectively. The best-fit photon indices are $2.1\pm0.1$ for NGC~424 and $2.0\pm0.1$ for NGC~1320, while for IC~2560 the fit runs into the upper domain limit of \mytorus at $\Gamma=2.6$. The fits are slightly better for higher assumed $N_{\rm H}$, but they never reach $\chi^2/\nu<1.5$. Most of the $\chi^2$ contribution comes from the iron line region, which we treat in more detail in \S\,\ref{sec:modeling3}. These results can easily be understood as the tendency of the fit to maximize the contribution of the reflected component (which increases with decreasing inclination, as the observer sees more of the inner far side of the torus), while not completely uncovering the source of the continuum at the center (since in the line of sight the light suffers significant absorption by passing through the edge of the torus for any inclination greater than $60^{\circ}$). In all cases the spectra are dominated by the reflection component, with only a minor contribution from transmission of the nuclear continuum along the line of sight.

Finally, we try a model in which the two spectral components of the \mytorus model are treated independently.\footnote{This is the {\em de-coupled} mode, after \citet{yaqoob-2012}.} In this case, the transmitted zeroth-order continuum, {\tt MYTZ}, and the scattered/reflection component, {\tt MYTS}, have fixed inclination parameters (90$^{\circ}$ for the former and 0$^{\circ}$ for the latter), separate column densities ($N_{\rm H,A}$ and $N_{\rm H,R}$, respectively) and a relative normalization ($A_{\rm rel}$) different than unity. This again leads to solutions in which the reflection component dominates over the transmission component (by a factor of $A_{\rm rel}=5-20$, compared to the internal normalization of the complete \mytorus model). However, the fit parameters are different for each AGN, as shown in the lower panels of Figure~\ref{fig:ss2}. For NGC~424 we find that a transmission component with $N_{\rm H,A}=(3\pm1)\times10^{23}$~cm$^{-2}$ contributes significantly in the iron line region, while the reflection continuum dominates above 10~keV. In the case of NGC~1320 the normalization of the transmitted component is consistent with zero. The lower quality of the IC~2560 spectum allows for a number of degenerate solutions that sensitively depend on the choice of assumptions. Assuming $\Gamma=2.55$ as before, one interesting possible solution ($\chi^2/\nu=27/30$) is reflection from Compton-thin material ($N_{\rm H,R}\approx5\times10^{23}$~cm$^{-2}$) dominating below 10~keV with a direct power law absorbed by $N_{\rm H,A}\approx5\times10^{24}$~cm$^{-2}$ dominating above that. Removal of the latter component, however, leads to a slightly better reflection-only model ($\chi^2/\nu=25/30$), which we elaborate on in \S\,\ref{sec:modeling3}. In conclusion, the physically motivated models of the AGN torus, in addition to the phenomenological models presented in \S\,\ref{sec:modeling1}, demonstrate that the observed \nustar spectra are consistent with being reflection-dominated.

\begin{figure*} 
\begin{center} 
\includegraphics[width=1.95\columnwidth]{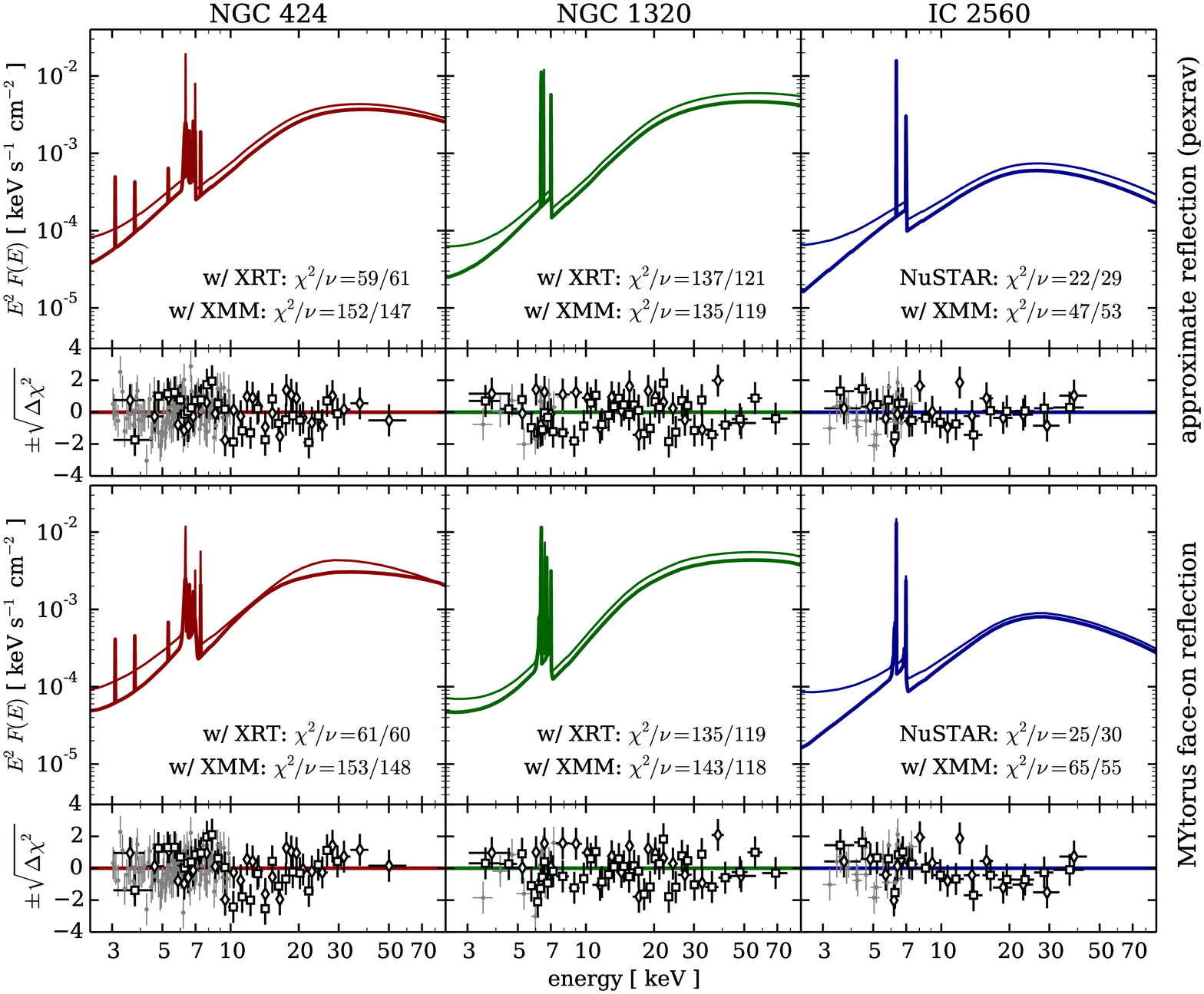}
\caption{ Reflection-only models fitted to the NGC~424, NGC~1320 and IC~2560 \nustar data jointly with simultaneous \swiftxrt and non-simultaneous archival \xmmnewton data (see \S\,\ref{sec:modeling3} for details). Thin colored lines show best fits to the \nustar and \swiftxrt data (except for IC~2560, where only the \nustar data was used), while the thick lines show the same for the \nustar and \xmm data lowered by 20\% for clarity. $\chi^2$ values for the best fits are given in each panel. Smaller panels show the residuals: black empty symbols for \nustar (diamonds for FPMA and squares for FPMB), and grey filled symbols for \xmm$\!\!$.  {\it Upper panels:} Models with the reflection continuum approximated by the \pexrav component. {\it Lower panels:} Models in which the reflection spectrum is represented by the face-on component of \mytorus$\!\!$. }
\label{fig:reflection_only}
\end{center} 
\end{figure*} 

\begin{figure*} 
\begin{center}
\includegraphics[width=2.0\columnwidth]{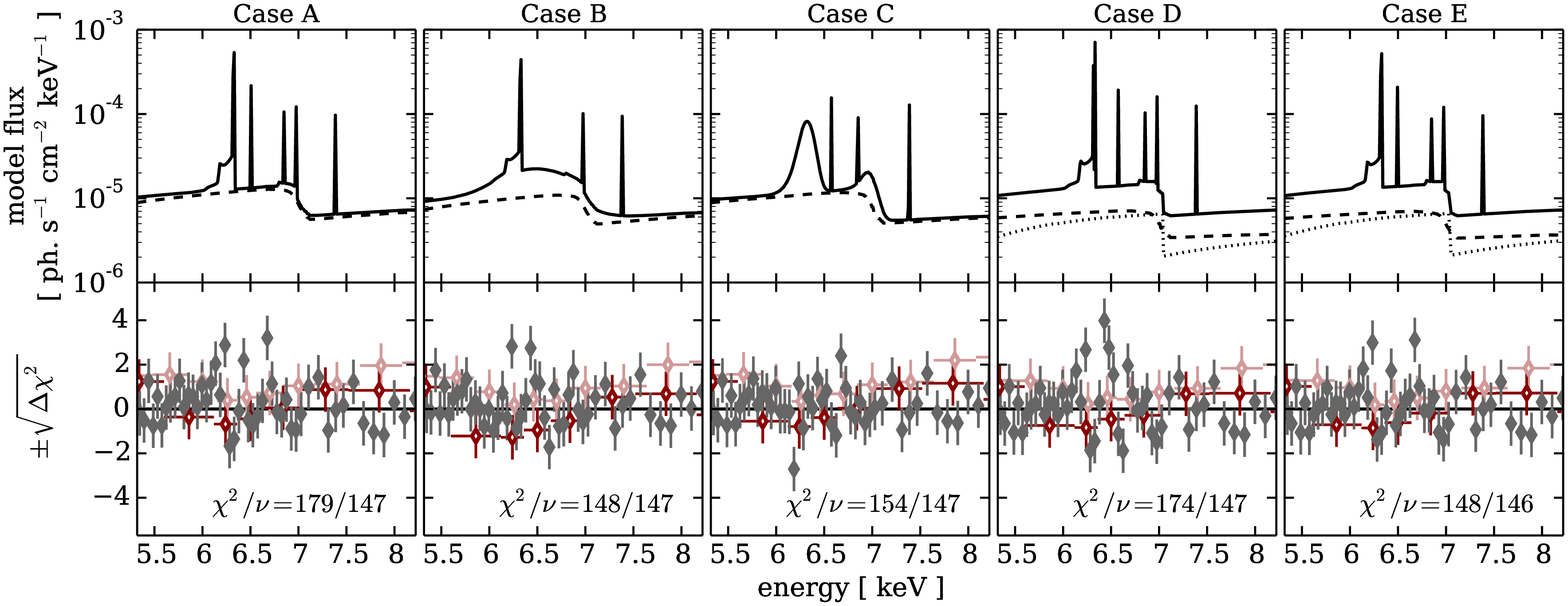}
\caption{ Demonstration of the spectral modeling of NGC~424 data in the energy range containing prominent iron lines. Different columns show the model curves (upper panel) and \xmm and \nustar residuals (lower panel) for specific cases discussed in \S\,\ref{sec:modeling3-ngc424}. The solid black lines in the upper panel show the sum of all model components, dashed lines show the reflection continuum, and dotted lines show the absorbed/transmitted component in the rightmost two panels. In the lower panels we show the \nustar FPMA (FPMB) residuals with dark (light) red symbols, and \xmm EPIC-pn residuals with grey symbols. In Case~A we fit for the energy of the ionized iron line at $\approx6.6$~keV, keeping energies of all other lines fixed; in Case~B, we let the width of that line to vary as well. In Case~C we broaden the Fe~K$\alpha$ and K$\beta$ lines while keeping line energies fixed at their expected values. Table~\ref{tab:model_parameters} lists the model parameters for Case~C. In Cases~D and~E we add a transmitted component with $N_{\rm H,A}\approx7\times10^{23}$~cm$^{-2}$, and fit for the energy of the ionized iron line in the latter. }
\label{fig:ngc424_lines}
\end{center}
\end{figure*} 

\subsection{Reflection-dominated Models} 

\label{sec:modeling3}

The conclusion of both of the preceeding two subsections is that the reflection-dominated models provide either better, or statistically equivalent but simpler, descriptions of the observed hard X-ray spectra compared to transmission-dominated or two-component models. Here we summarize the results obtained with the simultaneous \swift and \nustar data, and also consider the higher-quality, non-simultaneous, archival \xmmnewton data. In order to avoid the complexities associated with the detailed modeling of the soft X-ray emission unrelated to the AGN, we only use the \xmm data above 3~keV. The model parameters of interest are listed in Table~\ref{tab:model_parameters}.

\subsubsection{NGC~424} 

\label{sec:modeling3-ngc424}

Before using the \xmm data from \citet{marinucci+2011} jointly with the \nustar data, we first checked whether the target changed dramatically in flux between the two observations. We construct a simple phenomenological model for the 3--10~keV region by fitting the \xmm spectrum with a \pexrav continuum and Gaussian components for 8 emission lines, fixed to the following energies:\footnote{This is a somewhat simpler model than the one used in the original analysis, but it describes the 3--10~keV spectrum very well, with $\chi^2/\nu=0.95$. For identification of the various emission lines, see~\citet{marinucci+2011}.} 3.13, 3.83, 5.37, 6.40, 6.65, 6.93, 7.06 and 7.47~keV. None of the lines are resolved by \xmm$\!\!\!$, except the iron K$\alpha$ line at 6.4~keV with width of $\sigma=0.09\pm0.01$ keV. The 3--10~keV flux calculated for this model fitted to the \xmm data is $8.4\pm0.2\times10^{-13}$~\cgsflux, which is $\lesssim20$\% lower than the $1.1\pm0.1\times10^{-12}$~\cgsflux derived for the quasi-simultaneous \swiftxrt and \nustar$\!\!$/FPMA data fitted with the same model (all parameters fixed, except for the overall normalization factor). Given the cross-calibration uncertainty between \nustar and \xmm of 10\% (Madsen \etal$\!\!$, {\it in prep.}), the fluxes can be considered almost consistent. We therefore conclude that the flux variability is not severe and proceed with a joint spectral analysis.

The best-fit approximate reflection-only model (using \pexrav$\!\!$) for the NGC~424 data from \swift and \nustar is found for $\Gamma=1.71\pm0.09$, with $\chi^2/\nu=59/61$. The \xmm data require the Fe~K$\alpha$ line to be broadened ($\sigma=0.09\pm0.01$~keV, $\Delta\chi^2=109$ for one additional free parameter), but lead to a very similar result: $\Gamma=1.64\pm0.09$ with $\chi^2/\nu=152/147$. Strong Fe~K$\alpha$ lines with equivalent width of $\approx$1~keV are found in both cases. The \mytorus model fits the \nustar and \swiftxrt data well ($\chi^2/\nu=61/60$) for $\Gamma=2.28_{-0.09}^{+0.03}$. The best fit is achieved for reflector column density ($N_{\rm H,R}$) at the upper limit of the range covered by \mytorus$\!\!$, $10^{25}$~cm$^{-2}$, with a 90\% confidence lower limit of $5\times10^{24}$~cm$^{-2}$. We find that the line component normalization is marginally lower than unity and that the data favor addition of a narrow Ni~K$\alpha$ line at 7.47~keV ($\Delta\chi^2=7$ for one additional free parameter). The best fit parameters and their uncertainties at the 90\% confidence level are listed in Table~\ref{tab:model_parameters}. Model curves and residuals are shown in Figure~\ref{fig:reflection_only}.

A straightforward fit of the reflection-only \mytorus model does not find a statistically acceptable solution ($\chi^2/\nu=195/148$) for the joint \nustar and \xmm data. The \xmm residuals point towards a disagreement in the region surrounding the prominent iron lines between 6 and 8~keV. We show this energy range in more detail in Figure~\ref{fig:ngc424_lines}, with several different modeling solutions. We first attempt fitting for the energy of the ionized iron line otherwise fixed at 6.65~keV: in this case (Case~A) the fit is improved to $\chi^2/\nu=179/147$ for $E=6.57_{-0.03}^{+0.01}$~keV. Letting the width of the line vary in the fit (Case~B) leads to $E=6.54\pm0.06$~keV and $\sigma=0.33\pm0.08$, removing the need for the previously included line at 6.93~keV. Although this is a statistically good model ($\chi^2/\nu=148/147$), it is difficult the interpret the broad Gaussian feature that it includes. We find an equally good fit ($\chi^2/\nu=153/147$, Case~C) by broadening the \mytorus line component, which includes neutral Fe~K$\alpha$ and K$\beta$ lines, using a Gaussian kernel with $\sigma=0.06\pm0.01$~keV. Letting ionized iron line energy vary does not significantly improve the fit. The best-fit photon index is $\Gamma=2.07_{-0.09}^{+0.11}$ and the reflector column density is well constrained to $N_{\rm H,R}=(3\pm1)\times10^{23}$~cm$^{-2}$. The model parameters listed in Table~\ref{tab:model_parameters} represent this particular case.

An alternative two-component model is suggested by our modeling in \S\,\ref{sec:modeling2}, as well as the literature \citep{iwasawa+2001,marinucci+2011}. We add a second \mytorus component to describe the intrinsic continuum contribution transmitted through the absorbing torus ({\tt MYTZ}$\times${\tt pow} in \xspec$\!\!$), since the sharp iron edge of an absorbed power-law component ($N_{\rm H,A}\approx1\times10^{24}$~cm$^{-2}$) could significantly contribute to the 6--8~keV line region. We find that with emission line energies kept fixed (Case~D) the best fit occurs at $\chi^2/\nu=174/147$. The model is improved ($\chi^2/\nu=148/146$) if we additionally let one of the ionized iron lines' energy to vary in the fit; the best-fit energy in that case (Case~E) is $6.57_{-0.05}^{+0.02}$~keV. The photon index found in Cases~D and~E, which include component transmitted through a column density of $N_{\rm H,A}=(7\pm3)\times10^{23}$~cm$^{-2}$, is consistent with the one found in Case~C. The reflector column density is at the upper boundary of the model at $10^{25}$~cm$^{-2}$ for Cases~D and~E, with a 90\% confidence lower limit of $5\times10^{24}$~cm$^{-2}$. In \S\,\ref{sec:discussion} we discuss the physical plausibility of the simple solutions proposed here, however, we stress that the details of the iron line region modeling are entirely driven by the high-quality \xmm data, which are not the focus of this paper. Although some contribution of a heavily absorbed component cannot be completely ruled out, all of the X-ray data considered here support the reflection-dominated spectrum hypothesis within the statistical uncertainties.

\subsubsection{NGC~1320} 

\label{sec:modeling3-ngc1320}

NGC~1320 was simultaneously observed with \nustar and \swift twice. As no significant differences are apparent between the two observations, we model both epochs simultaneously and list the best-fit parameters of those fits in Table~\ref{tab:model_parameters}. The best-fit photon index is hard ($\Gamma=1.3\pm0.1$), however, assuming a higher cut-off energy for the intrinsic continuum brings it closer to the typical value: e.g.,~for a cut-off at 500~keV, the best fit is obtained for $\Gamma=1.5\pm0.1$. As in the case of NGC~424, we find that in the \mytorus reflection-only model the best-fit photon index is steeper ($\Gamma=1.9_{-0.3}^{+0.1}$) than the photon index derived from the \pexrav modeling. The joint fit of the approximate reflection-only model with \pexrav is significantly improved upon adding a narrow line component at $6.57_{-0.08}^{+0.16}$~keV, which is most likely an Fe~XXV K$\alpha$ line. Using the \mytorus reflection-only model, a line is required at $6.6\pm0.1$~keV. In both cases, the equivalent width of the line is $0.3\pm0.2$~keV. All other spectral parameters are consistent between the two. Addition of the ionized iron line is essentially the only improvement needed over the reflection-dominated models already mentioned in \S\,\ref{sec:modeling1} and \S\,\ref{sec:modeling2}. We find that the normalization of the lines component of the \mytorus model is mildly elevated, but consistent with unity ($1.2_{-0.4}^{+0.3}$). The \mytorus model additionally provides a constraint on the column density of the reflecting material, instead of assuming it to be infinite. The reflector column density, $N_{\rm H,R}$, is mildly degenerate with the intrinsic photon index, but it can be constrained independently to $N_{\rm H,R}=(4_{-2}^{+4})\times10^{24}$~cm$^{-2}$ with 90\% confidence.

The only soft X-ray data available for NGC~1320 besides our quasi-simultaneous \swiftxrt data is from a relatively short 12-ks \xmmnewton observation in 2006 \citep{brightman+nandra-2011a}. Only a minor difference of $\lesssim30$\% in the 3--10~keV flux is observed between the \xmm and \swift$\!\!+$\nustar observations, so we proceed with a joint analysis. No significant difference is found between the best-fit intrinsic photon indices based on the \swift or \xmm data for either the \pexrav or \mytorus models. A significant improvement in either case is found if a narrow line corresponding to ionized iron is added to the model The best fit energy is $6.55_{-0.09}^{+0.10}$~keV for the \pexrav model and $E=6.77_{-0.23}^{+0.06}$~keV for the \mytorus model, with equivalent widths of $0.2\pm0.1$~keV and $0.3_{-0.2}^{+0.1}$~keV, respectively. Addition of a Ni~K$\alpha$ line at 7.47~keV does not significantly improve the fit. The \xmm data further constrain the column density of the reflector: $N_{\rm H,R}=(4_{-1}^{+2})\times10^{24}$~cm$^{-2}$ with 90\% confidence. Both best-fit models and their respective residuals are shown in the middle column of Figure~\ref{fig:reflection_only}.

\subsubsection{IC~2560} 

\label{sec:modeling3-ic2560}

IC~2560 is the faintest target considered in this paper, and correspondingly has the poorest photon statistics. The quasi-simultaneous \swiftxrt observation is too short to provide useful soft X-ray data, so for the initial modeling we use the \nustar data alone. From the best fit of the approximate \pexrav model we find that the intrinsic photon index is steeper than in the other two targets ($\Gamma=2.2_{-0.4}^{+0.3}$) and that the iron lines are strong; the equivalent widths of the Fe~K$\alpha$ and K$\beta$ lines are $2.1_{-0.5}^{+1.3}$~keV and $0.6_{-0.4}^{+0.6}$~keV, respectively. In the case of the \mytorus reflection-only model we find that the normalization of the line component is significantly elevated, $2.2_{-0.5}^{+0.7}$. The intrinsic photon index is somewhat degenerate with the reflector column density and both are best fitted by parameter values on the edge of the validity domain of the model. Formally, we are able to derive only the lower limits on the best-fit parameters: $\Gamma>2.6$ and $N_{\rm H,R}>10^{25}$~cm$^{-2}$. By fixing the photon index to 2.55 (which is statistically acceptable, with $\chi^2/\nu=25/30$), we can estimate a 90\% confidence lower limit on $N_{\rm H,R}$ to be 7$\times10^{24}$~cm$^{-2}$ based on the \nustar data alone.

The \xmm data we use are MOS1 and MOS2 spectra (above 3~keV) from an 80-ks observation published in~\citet{tilak+2008}. By calculating 3--10~keV fluxes based on a simple reflection model represented by a sum of a \pexrav continuum and a Gaussian line component we find that the flux did not change between the \xmm and \nustar observation by more than 20\%. The joint fits to the \nustar and \xmm data result in best-fit parameters entirely consistent with those found with the \nustar data alone. Again, the best-fit photon index and column density in the \mytorus model is formally outside of its validity domain. By assuming statistically acceptable $\Gamma=2.55$ ($\chi^2/\nu=65/55$) we can constrain $N_{\rm H,R}$ to be greater than 7$\times10^{24}$~cm$^{-2}$ with 90\% confidence. Both models and their residuals are shown in the rightmost column of Figure~\ref{fig:reflection_only}. Table~\ref{tab:model_parameters} provides the list of the best-fit values for all relevant model parameters.

\begin{deluxetable*}{r cc cc cc} 
\tabletypesize{\scriptsize}
\tablewidth{1.9\columnwidth}
\tablecolumns{7}

\tablecaption{ Summary of the reflection-only models fitted to the quasi-simultaneous \nustar and \swiftxrt data, and non-simultaneous \nustar and \xmmnewton data. See \S\,\ref{sec:modeling3} for details. Uncertainties listed here are 90\% confidence intervals. (f) marks fixed parameters. \label{tab:model_parameters} }

\tablehead{
  \colhead{} &
  \multicolumn{2}{c}{\bf NGC~424} &
  \multicolumn{2}{c}{\bf NGC~1320} &
  \multicolumn{2}{c}{\bf IC~2560} \\
  \cline{2-3} \cline{4-5} \cline{6-7} \\
  \colhead{Data Used} &
  \colhead{\nustar+} &
  \colhead{\nustar+} &
  \colhead{\nustar+} &
  \colhead{\nustar+} &
  \colhead{\multirow{2}{*}{\nustar}} &
  \colhead{\nustar+} \\
  \colhead{for the Fit} &
  \colhead{\swiftxrt} &
  \colhead{\xmmnewton\tablenotemark{a}} &
  \colhead{\swiftxrt} &
  \colhead{\xmmnewton\tablenotemark{a}} &
  \colhead{} &
  \colhead{\xmmnewton\tablenotemark{b}} \\
}

\startdata

\cutinhead{ approximate reflection model, {\tt C}\,$\times$\,{\tt phabs( pow $+$ pexrav $+$ zgauss[}$\times 5${\tt ] )}}
$\chi^2/\nu$								& 59/61 & 152/147 & 137/121 & 135/119 & 22/29 & 47/54 \\
$\Gamma$ (\,\pexrav$\!\!$)\,\tablenotemark{c}	& $1.71\pm0.09$ & $1.66\pm0.09$ & $1.3\pm0.1$ & $1.3\pm0.1$ & $2.2_{-0.4}^{+0.3}$ & $2.2_{-0.2}^{+0.1}$ \\
$EW_{\mbox{\tiny Fe K$\alpha$}}$ [ keV ]		& $1.0\pm0.3$ & $1.07_{-0.09}^{+0.13}$\,\tablenotemark{d} & $0.7_{-0.5}^{+1.7}$ & $1.1_{-0.4}^{+0.9}$ & $2.1_{-0.5}^{+1.3}$ & $2.5_{-0.4}^{+0.6}$ \\
$EW_{\mbox{\tiny Fe K$\beta$}}$ [ keV ]		& $0.4_{-0.2}^{+0.3}$ & $0.13\pm0.04$ & $0.5\pm0.3$ & $0.6_{-0.4}^{+0.6}$ & $0.8\pm0.6$ & $0.7\pm0.3$ \\
$EW_{\mbox{\tiny Ni K$\alpha$}}$ [ keV ]		& \nodata & $0.16\pm0.06$ & \nodata & \nodata & \nodata & \nodata \\
$EW_{\mbox{\tiny ion.$\!$ Fe}}$ [ keV ]		& \nodata & $0.05\pm0.02$, $0.10_{-0.04}^{+0.05}$ & $0.3\pm0.2$ & $0.2\pm0.1$ & \nodata & \nodata \\
$E_{\mbox{\tiny ion.$\!$ Fe}}$ [ keV ]		& \nodata & 6.65 (f), 6.93 (f) & $6.57_{-0.08}^{+0.16}$ & $6.54_{-0.09}^{+0.10}$ & \nodata & \nodata \\
{\tt C\ }$(\mbox{\tiny FPMB/FPMA})$			& $1.1\pm0.1$ & $1.1\pm0.1$ & $0.93\pm0.07$ & $0.9\pm0.1$ & $1.1\pm0.2$ & $1.1_{-0.2}^{+0.1}$ \\
{\tt C\ }$(\mbox{\tiny SOFT/FPMA})$			& $1.4\pm0.4$ & $1.1\pm0.1$ & $1.1\pm0.3$ & $0.8\pm0.1$ & \nodata & $0.9_{-0.1}^{+0.2}$ \\

\cutinhead{ \mytorus face-on reflection model, {\tt C}\,$\times$\,{\tt phabs( pow $+$ MYTS[}$i=0${\tt ] $+$ K}\,$\times$\,{\tt MYTL[}$i=0${\tt ] $+$ zgauss[}$\times 3${\tt ] )}}
$\chi^2/\nu$							& 61/60 & 153/148\,\tablenotemark{e} & 135/121 & 143/118 & 25/30 & 65/55 \\
$\Gamma$ (\,\mytorus$\!\!$)				& $2.28_{-0.09}^{+0.03}$ & $2.07_{-0.09}^{+0.11}$ & $1.7\pm0.2$ & $1.6\pm0.2$ & 2.55 (f) & 2.55 (f) \\
$N_{\rm H,R} \left[10^{24}\ \mbox{cm}^{-2}\right]$\,\tablenotemark{f} & $10_{-5}^{+u}$ & $3\pm1$ & $4_{-2}^{+4}$ & $4_{-1}^{+2}$ & $10_{-3}^{+u}$ & $10_{-3}^{+u}$ \\
line norm. (\,{\tt K}\,)\,\tablenotemark{g}		& $0.7\pm0.2$ & $0.68_{-0.07}^{+0.09}$ & $1.2_{-0.4}^{+0.3}$ & $1.6\pm0.2$ & $2.2_{-0.5}^{+0.7}$ & $2.1\pm0.3$ \\
$EW_{\mbox{\tiny Ni K$\alpha$}}$ [ keV ]	& $0.5\pm0.4$ & $0.04\pm0.01$ & \nodata & \nodata & \nodata & \nodata \\
$EW_{\mbox{\tiny ion.$\!$ Fe}}$ [ keV ]	& \nodata & $0.04\pm0.02$, $0.05\pm0.03$ & $0.3\pm0.2$ & $0.3_{-0.2}^{+0.1}$ & \nodata & \nodata \\
$E_{\mbox{\tiny ion.$\!$ Fe}}$ [ keV ]	& \nodata & 6.65 (f), 6.93 (f) & $6.6\pm0.1$ & $6.77_{-0.23}^{+0.06}$ & \nodata & \nodata \\
{\tt C\ }$(\mbox{\tiny FPMB/FPMA})$		& $1.1\pm0.1$ & $1.1\pm0.1$ & $0.93\pm0.08$ & $0.93\pm0.07$ & $1.1\pm0.2$ & $1.1\pm0.2$ \\
{\tt C\ }$(\mbox{\tiny SOFT/FPMA})$		& $1.4\pm0.4$ & $1.1\pm0.1$ & $0.9_{-0.3}^{+0.4}$ & $0.8\pm0.1$ & \nodata & $1.0\pm0.2$ \\

\enddata

\tablenotetext{a}{EPIC/pn data, used only above 3~keV.} 
\tablenotetext{b}{MOS1 and MOS2 data, used only above 3~keV. The cross-normalization with \nustar$\!\!$/FPMA given in the table is the mean between the two.} 
\tablenotetext{c}{Set to produce only the reflection continuum (i.e. no contribution from the intrinsic power law continuum).} 
\tablenotetext{d}{Iron K$\alpha$ line width was determined by fitting, $\sigma=0.09\pm0.01$~keV.} 
\tablenotetext{e}{In this fit the line component is Gaussian-smoothed with $\sigma=0.06\pm0.01$~keV. Note that simple alternative models can be found as well, as discussed in \S\,\ref{sec:modeling3-ngc424}.} 
\tablenotetext{f}{Hydrogen column density of the material producing the reflection spectrum. Values with uncertainty marked with $+u$ denote $90\%$ confidence upper limits in excess of $10^{25}$~cm$^{-2}$, which is outside of the domain of the \mytorus model.} 
\tablenotetext{g}{Relative normalization of the emission lines component with respect to its corresponding reflection continuum. \vspace{0.5cm}} 

\end{deluxetable*} 

\begin{figure*} 
\begin{center}
\includegraphics[width=1.95\columnwidth]{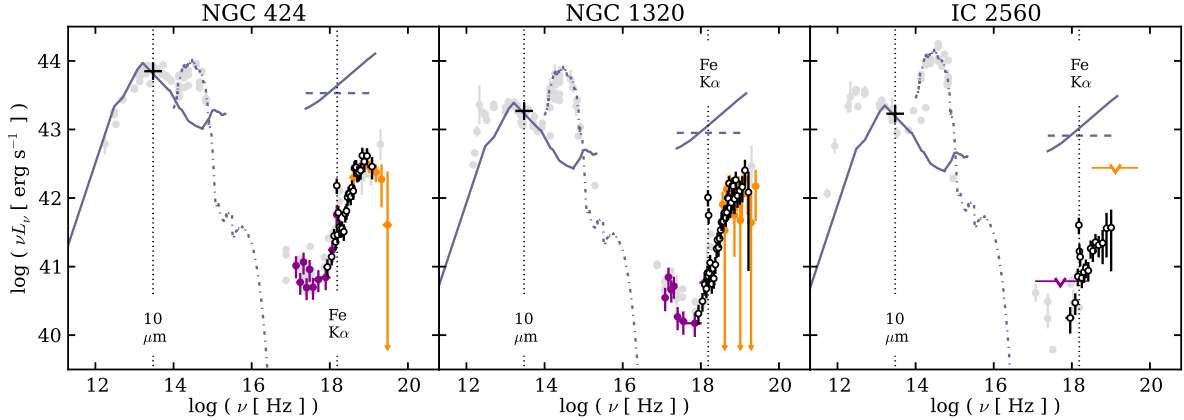}
\caption{ SEDs of NGC~424 (left), NGC~1320 (middle), and IC~2560 (right panel) constructed using archival data (grey points) acquired using the ASDC SED Tool. The minor vertical spread in the data in the optical and infrared, only reflects the fact that different finite apertures had been used for photometry in different publicly available catalogs. The black crosses mark the luminosities at 10~$\mu$m, considered as representative mid-infrared luminosities ($L_{\rm MIR}$) in this work. The frequency of the prominent neutral Fe~K$\alpha$ line at 6.4~keV is marked in all panels. We show an Sa/b galaxy template with a dark grey line as a proxy for the host galaxy SED. The \swift spectra are shown in magenta and orage symbols for the XRT and the BAT instruments, respectively. For IC~2560 only upper limits are available (marked with empty downward arrows; see text for details). The errorbars on some \swiftbat bins extend below the the y-axis limit, and are marked with small filled downward arrows. The \nustar data are plotted in black. }
\label{fig:seds}
\end{center}
\end{figure*} 

\section{Discussion} 

\label{sec:discussion}

\subsection{Comparison with Previously Published\\ X-ray Spectral Analyses} 

\label{sec:discussion-previous}

The earliest X-ray spectrum of NGC~424 came from an {\em ASCA} observation, which revealed a prominent iron line and a hard spectrum suggestive of Compton-thick reflection of the nuclear continuum \citep{collinge+brandt-2000}. This result has been confirmed in later observations, by \bepposax$\!\!$, \chandra and \xmmnewton \citep{iwasawa+2001,matt+2003a}. The soft X-ray spectrum has most recently been analysed in depth by \citet{marinucci+2011}, using a long $\sim$100-ks \xmmnewton observation. The focus of that work was on detailed modeling of the physical state of the plasma dominating below 2~keV, but the data were also used to model the X-ray continuum and line emission up to 10~keV. In agreement with earlier results, they found support for a strong reflection component and a heavily absorbed power-law continuum obscured by nearly Compton-thick material with $N_{\rm H,A}=1.1\times10^{24}$~cm$^{-2}$ contributing only above 5~keV. With the hard X-ray coverage of \bepposax \citep{iwasawa+2001} it was possible to infer that the intrinsic power-law continuum is absorbed by a column density of $\gtrsim2\times10^{24}$~cm$^{-2}$. Note, however, that only the simple approximate models were used in the spectral analyses leading to the inference of the column density, and that they required assuming a photon index ($\Gamma=2$), since it was not possible to constrain it directly from the data. Using the simultaneous \swiftxrt and \nustar data, we firmly establish that the hard X-ray spectrum of NGC~424 can be described as being dominated by reflection. A contribution from a heavily absorbed component cannot be completely ruled out, but it is not formally required by any of the data considered in this work.

\citet{burlon+2011} analyzed the first three years of \swiftbat data on NGC~424 and modeled it simply as a heavily absorbed power law with $N_{\rm H,A}=(2.0_{-0.4}^{+0.3})\times10^{24}$~cm$^{-2}$ and $\Gamma=1.9\pm0.3$. Since that model does not include any contribution from a reflection component, it is not directly comparable to our results. The \swiftbat spectrum from the 70-month survey \citep{baumgartner+2013-swiftbat70} is entirely consistent with the \nustar and \xmm data ($\chi^2/\nu=153/153$; see Figure~\ref{fig:seds}) for a cross-normalization constant of $1.5\pm0.3$ relative to \nustar$\!\!$/FPMA. We also verify that our models are consistent with the \bepposax data from \citet{iwasawa+2001}. With the photon index fixed at its best-fit value for that model, $\Gamma=1.68$, we find that the cutoff energy is $E_{\rm cut}=120_{-30}^{+50}$~keV. A slightly higher cut-off energy, $E_{\rm cut}=190_{-80}^{+260}$~keV, is inferred from the equivalent fit using the \swiftxrt instead of the \xmm data. The 14--195~keV luminosity published in the 70-month \swiftbat catalog ($6.5\times10^{42}$~erg~s$^{-1}$) is calculated by assuming a relatively flat $\Gamma\approx2$ spectrum; if we use our model instead, the 14--195~keV luminosity based on \swiftbat data alone is $5.5\times10^{42}$~erg~s$^{-1}$. The apparently significant normalization offset and its uncertainty are likely due to the limited statistics and long-term averaging of the \swiftbat data, in addition to simple flux calibration differences. In the rest of the discussion we assume $\sim30$\% lower luminosity as inferred from the \nustar data, as listed in Table~\ref{tab:luminosities}.

No dedicated long X-ray observations of NGC~1320 exist in the literature: the only previously available soft X-ray data was taken with \xmmnewton as part of a recent survey of infrared-bright AGN \citep{brightman+nandra-2011a}. The original modeling by those authors and later re-analyses (e.g.,\,\citealt{georgantopoulos+2011,severgnini+2012,marinucci+2012}) agree that the nucleus of NGC~1320 is heavily obscured ($N_{\rm H,A}\gtrsim1\times10^{24}$~cm$^{-2}$) and infer the presence of a considerable reflection component on the basis of a strong iron line with an equivalent width of $\sim$1~keV. \citet{gilli+2010} assert that both reflected and transmitted components contribute to the spectrum, but provide very few details on the modeling as the model parameters are largely unconstrained by the data. Our analysis (which includes the same \xmm data in addition to the \nustar data) confirms most of the earlier results and solidifies the dominance of the reflection spectrum above 2~keV. A transmitted component is not formally required by our data. A hard X-ray source was detected by \swiftbat$\!\!$ at the coordinates of NGC~1320, but the low significance of that detection does not provide any additional spectral constraints \citep{cusumano+2010-swiftbat54}. The published 15--150~keV luminosity, $2.8\times10^{42}$~erg~s$^{-1}$ (observed, uncorrected for absorption), agrees well with the luminosity calculated from our spectral model in the slightly wider 14--195~keV band (Table~\ref{tab:luminosities}).

Prior X-ray observations of IC~2560 have been performed by \asca \citep{risaliti+1999}, \chandra \citep{iwasawa+2002,madejski+2006} and \xmmnewton \citep{tilak+2008}. The earliest observation already showed that the source was likely Compton-thick, with a very strong iron line indicating the presence of a reflection component. More sensitive observations with \chandra and later \xmm confirmed that the 2--10~keV spectrum can be well described as a cold neutral reflection from a Compton-thick medium, including a particularly strong iron line with an equivalent width in excess of 2~keV. The lack of a heavily absorbed intrinsic continuum component in all observations has been explained by invoking Compton-thick nuclear obscuration. The most stringent constraint on the absorption column density is provided by a 80-ks \xmm observation: $N_{\rm H,A}>3\times10^{24}$~cm$^{-2}$ \citep{tilak+2008}. This source is very faint at hard X-ray energies, as confirmed by our \nustar data, and has never previously been detected by any hard X-ray instrument above 10~keV. This is partly due to the heavy obscuration, but possibly also due to the atypically steep intrinsic photon index ($\Gamma>2.2$) suggested by our modeling. Owing to the lack of high-energy coverage, previous studies could not constrain the photon index. The lack of detection of a heavily absorbed transmitted component in the \nustar band satisfies all previous lower limits on the line-of-sight column density, and pushes it further towards the $\sim10^{25}$~cm$^{-2}$ regime.

\subsection{Multi-wavelength Data\\ and Spectral Energy Distributions} 

\label{sec:discussion-seds}

All three of our sources have been previously observed in a wide range of spectral bands. We use publicly available archival data from the ASDC SED Tool\footnote{\texttt{http://tools.asdc.asi.it/SED/}} to construct the rough spectral energy distributions shown in Figure~\ref{fig:seds}. Note that the distance uncertainties translate into a systematic uncertainty of approximately 0.1 dex in the vertical direction. As it is not important for our work, we do not concern ourselves with the various aperture diameters allowing differing levels of host stellar contamination in the optical and near-infrared photometry. We filter out small-aperture measurements, and those with large and unspecified uncertainties. All three galaxies are morphologically classified to be at a transition between S0 and Sa/b type \citep{devaucouleurs+1991}. The starlight dominates the optical output (the nuclei are heavily extincted in the optical), but most of the mid-infrared luminosity can be ascribed to the AGN (as thermal radiation from the torus), since there are no indications of significant starburst activity in any of the sources. By averaging over the wealth of mid-infrared data available for each of the sources, we estimate their mid-infrared luminosities ($L_{\rm MIR} = \nu L_{\nu}$ at $\sim$10~$\mu$m) to be $7\times10^{43}$~erg~s$^{-1}$ for NGC~424 and $2\times10^{43}$~erg~s$^{-1}$ for both NGC~1320 and IC~2560 (see Table~\ref{tab:luminosities}). These values are expected to be different by no more than a factor of $\sim$2 from the luminosities in any of the mid-infrared bands between 5 and 25~$\mu$m commonly used in the literature, which is acceptable for the purposes of our order-of-magnitude calculations in \S\,\ref{sec:discussion-luminosities}.

The quasi-simultaneous \swift and \nustar data for all sources are plotted together with the multi-wavelength archival data in Figure~\ref{fig:seds}. The short 2-ks \swiftxrt observation of IC~2560 did not provide useful data for spectral modeling, so we show only the $3\sigma$ upper limit on the 0.3--10~keV flux of $\sim2\times10^{-13}$~erg~s$^{-1}$~cm$^{-2}$. This corresponds to the luminosity upper limit plotted as a downward arrow in the soft X-ray range of Figure~\ref{fig:seds} (rightmost panel). The \swiftxrt and \nustar data for NGC~424 and NGC~1320, as well as the \nustar data for IC~2560, were unfolded in \xspec using their respective best-fit models. The same models were used to unfold the \swiftbat spectra, since they fit those data well. The highest-quality \swiftbat spectrum of NGC~424 was taken from the 70-month catalog~\citep{baumgartner+2013-swiftbat70}. The NGC~1320 data is only available as a part of the 54-month catalog~\citep{cusumano+2010-swiftbat54}, since it is undetected in the 70-month catalog due to low statistics and different metodology.

IC~2560 is not listed in any of the published \swiftbat source catalogs. The detection threshold of the 70-month catalog at 5$\sigma$ can be calculated from Equation~(9) in \citet{baumgartner+2013-swiftbat70} using an estimate of the exposure time in the part of the sky surrounding IC~2560. Based on their Figure~1, we estimate the exposure time to be approximately 12~Ms and derive a sensitivity of $9\times10^{-12}$~erg~s$^{-1}$~cm$^{-2}$ in the 14--195~keV band. This upper limit is displayed as an orange downward arrow in Figure~\ref{fig:seds}. The highest-energy \nustar data points are almost an order of magnitude below this limit. 

All three sources are relatively faint in the radio, with none of their radio luminosities ($L_{\rm R}=\nu L_{\nu}$, for $\nu=5$~GHz) exceeding $2\times10^{38}$~erg~s$^{-1}$, based on archival flux densities (see Table~\ref{tab:luminosities}). \citet{terashima+wilson-2003} defined that radio-loud AGN have $R_{\rm X}>-4.5$, where $R_{\rm X}=\log \left( L_{\rm R} / L_{\rm X}\right)$, $L_{\rm R}$ is as defined above, and $L_{\rm X}$ is the intrinsic 2--10~keV luminosity. Despite the latter being uncertain due to obscuration, as we elaborate in the following section, all three AGN may at best straddle the dividing line between the radio-quiet and radio-loud objects, but they are not strong radio sources. Using the empirical relations of \citet{bell-2003} to convert the radio luminosity into star formation rate, we find $\sim$1~$M_{\odot}$~yr$^{-1}$ in each case. The radio luminosity is thus consistent with being due to star formation typical for the S0--Sa/b morphology of the host galaxies \citep{bendo+2002}.

\begin{deluxetable}{r c c c}

\tabletypesize{\scriptsize}
\tablecaption{ Estimated luminosities of our targets in different spectral bands in units of $10^{41}$~erg~s$^{-1}$. All luminosities are observed, i.e.~they are not corrected for absorption or reddening. \label{tab:luminosities} }

\tablehead{
  \colhead{ } &
  \colhead{\multirow{2}{*}{\bf NGC~424}} &
  \colhead{\multirow{2}{*}{\bf NGC~1320}} &
  \colhead{\multirow{2}{*}{\bf IC~2560}} \\
  \colhead{ } &
  \colhead{ } &
  \colhead{ } &
  \colhead{ } }

\startdata

14--195~keV\tablenotemark{a} & 36 & 27 & 5.0 \\
15--55~keV\tablenotemark{a} & 23 & 14 & 3.6 \\
2--10~keV\tablenotemark{a} & 2.5 & 0.90 & 0.80 \\
$[$O\,III$]$ $\lambda$5007\tablenotemark{b} & 1.7 & 0.20 & 0.25 \\
MIR (10~$\mu$m)\tablenotemark{b} & 710 & 190 & 170 \\
5~GHz\tablenotemark{c} & 0.0018 & 0.0003 & 0.0016 \\

\enddata

\tablenotetext{a}{Calculated from our best-fit models. Typical statistical uncertainty is $\lesssim20$\%.}
\tablenotetext{b}{Average value based on published measurements in \citet{murayama+1998,gu+2006,lamassa+2010,kraemer+2011}. Overall uncertainties in absolute values are estimated to be about a factor of two. }
\tablenotetext{c}{Calculated from NVSS flux density at 1.4~GHz \citep{condon+1998}, assuming a $f_{\nu} \propto \nu^{-0.7}$ spectrum. \vspace{0.5cm}}

\end{deluxetable}

\subsection{Intrinsic X-ray Luminosities} 

\label{sec:discussion-luminosities}

The three AGN considered in this work are all heavily obscured, as previously suggested by their relatively faint soft X-ray spectra with strong fluorescent iron lines, and strong mid-infrared and [O\,III] emission (e.g.,~\citealt{risaliti+1999,collinge+brandt-2000,brightman+nandra-2011b}). The \nustar data presented here confirm that their intrinsic continua are heavily suppressed by obscuration and thus not directly observable. For a column density of the order of $10^{24}$~cm$^{-2}$ (and particular obscuring material geometry) the intrinsic X-ray continua would significantly contribute to the flux in the \nustar 3--79~keV band despite the heavy obscuration. However, our modeling suggests that the hard X-ray spectra of NGC~424, NGC~1320 and IC~2560 are dominated by a reflected component in each case, indicating that along the line of sight towards those nuclei the absorption column is well above the Compton-thick threshold at $1.4\times10^{24}$~cm$^{-2}$ and likely in the $10^{25}$~cm$^{-2}$ regime.

Without a direct constraint on the transmission of the intrinsic continuum along the line of sight, we can only indirectly infer the intrinsic X-ray luminosity through other SED components. NGC~1320 and IC~2560 have very similar mid-infrared, [O\,III] and 2--10~keV luminosities (see Table~\ref{tab:luminosities}). We note that star formation may be contributing a part of the luminosity in these bands, but we ignore it in the simple estimates performed here. Comparing to empirical correlations from the literature, we find that without any correction for absorption NGC~1320 and IC~2560 fall at least an order of magnitude below the distribution of intrinsic $L_{\rm 2-10~keV}/L_{\rm MIR}$ and $L_{\rm 2-10~keV}/L_{\rm [O\,III]}$ ratios. For example, for NGC~1320 (IC~2560) $\log(L_{\rm 2-10~keV}/L_{\rm MIR})=-2.3$ ($-2.4$) compared to $-0.63\pm0.69$ \citep{lutz+2004}, and $\log(L_{\rm 2-10~keV}/L_{\rm [O\,III]})=0.6$ (0.5) compared to $1.76\pm0.38$ \citep{mulchaey+1994}. The quoted median log-ratios and their standard deviations were derived empirically from sample studies of nearby Seyfert~2 nuclei. Simply assuming that these intrinsic ratios hold for our AGN leads to a conservative lower limit on the intrinsic 2--10~keV luminosity of $>2\times10^{42}$~erg~s$^{-1}$.

\begin{figure} 
\begin{center}
\includegraphics[width=0.98\columnwidth]{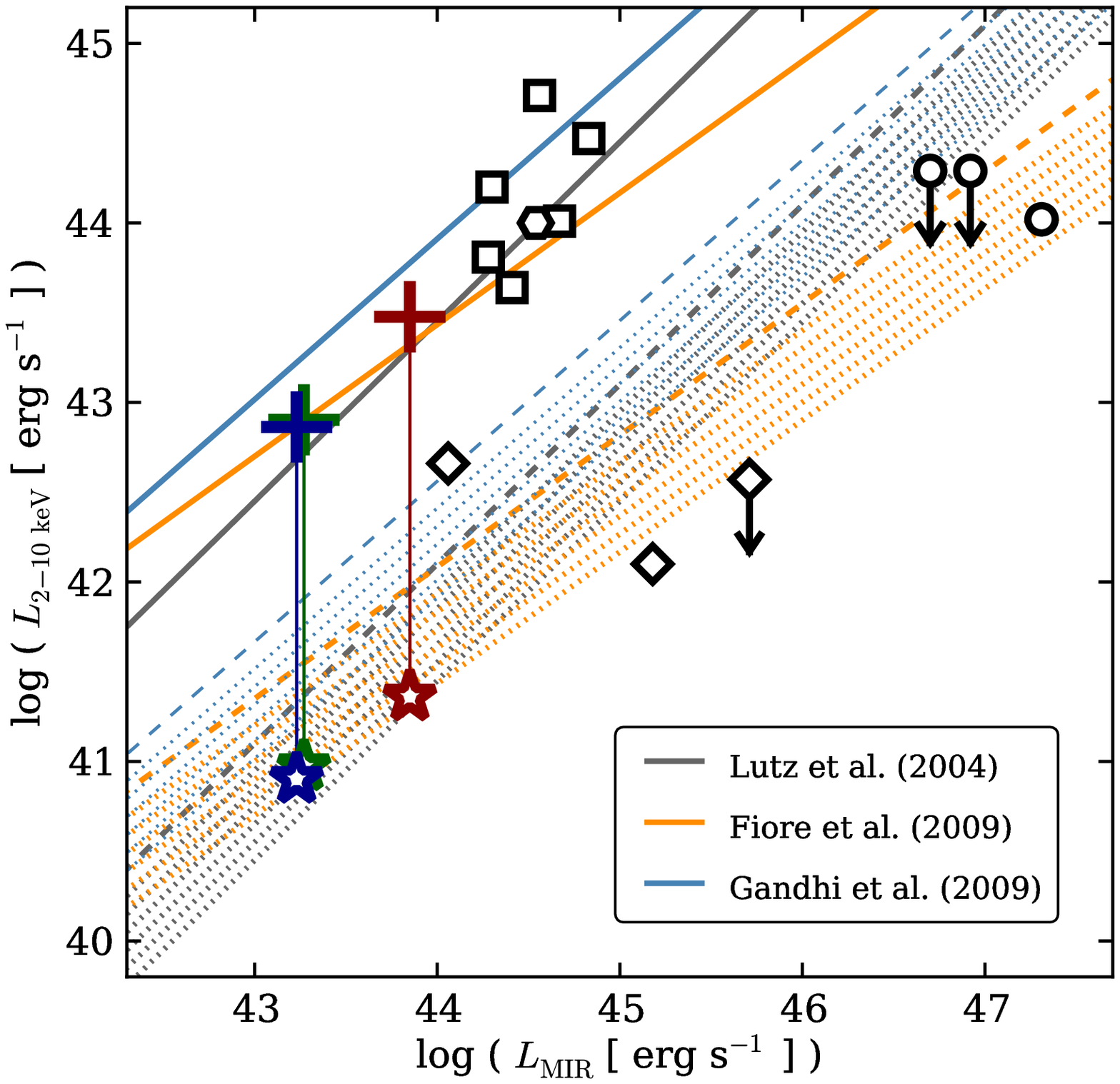}
\caption{ The solid lines show empirical $L_{\rm 2-10~keV}-L_{\rm MIR}$ correlations from the literature: \citet{lutz+2004} in dark grey, \citet{fiore+2009} in yellow, and \citet{gandhi+2009} in light blue. With dashed lines we show the same relations for sources absorbed by a column density $N_{\rm H,A}=1.4\times10^{24}$~cm$^{-2}$, i.e. for barely Compton-thick absorption. The dotted lines mark regions occupied by reflection-dominated sources (i.e.,~the absorbed continuum component is not observable at all below 10~keV) with reflection efficiency between 1 and 3\%, calculated from the \mytorus model. NGC~424, NGC~1320 and IC~2560 are plotted in dark red, green and blue, respectively. Their observed 2--10~keV luminosities are marked with star symbols, while the crosses show the adopted intrinsic values (with their size approximating the uncertainty). Overplotted in empty black symbols we show other AGN observed with \nustar in the extragalactic surveys program: serendipitously discovered non-beamed AGN from \citet{alexander+2013} with squares, SDSS-selected obscured quasars from \citet{lansbury+2014} with diamonds, WISE-selected ultra-lumninous $z=2$ quasars from \citet{stern+2014} with circles, and a $z=2$ quasar identified in the ECDFS field from \citet{delmoro+2014} with a hexagon.}
\label{fig:luminosities}
\end{center}
\end{figure}

In the case of obscuration by a column density of $\lesssim1.4\times10^{24}$~cm$^{-2}$ the transmitted continuum in the 2--10~keV band is $\sim$10\% of the intrinsic flux (e.g.,~\citealt{burlon+2011}), but in order for the reflection to dominate the X-ray spectrum, a suppression by a factor of $\sim$100 is typically needed (e.g.,~\citealt{matt+1997,matt+2004}). We therefore assume that the intrinsic 2--10~keV luminosity of both sources is roughly a factor of 100 greater than the observed one, estimating it at $\lesssim1\times10^{43}$~erg~s$^{-1}$. With this luminosity and the mid-infrared luminosities of almost $2\times10^{43}$~erg~s$^{-1}$, NGC~1320 and IC~2560 fall right onto the tight relation between $L_{\rm 2-10~keV}$ and $L_{\rm MIR}$ from \citet{gandhi+2009}, where both quantities are intrinsic (i.e.\, the infrared luminosity directly measures the torus emission) and the relation is directly applicable to heavily obscured AGN. For comparison, \citet{tilak+2008} derive $\sim6\times10^{42}$~erg~s$^{-1}$ from the mid-infrared luminosity of IC~2560, however, they also arrive at an order of magnitude lower estimate based on the [O\,III] luminosity\footnote{Note that earlier estimates of the intrinsic luminosity \citep{ishihara+2001,iwasawa+2002,madejski+2006} generally yielded lower values, which is partly due to significantly lower assumed distance to IC~2560: 26~Mpc, compared to 41.4~Mpc used in this work.}. For NGC~1320 \citet{brightman+nandra-2011a} derive an intrinsic 2--10~keV luminosity of $5\times10^{42}$~erg~s$^{-1}$. Our best estimate is therefore $7\times10^{42}$~erg~s$^{-1}$ for both sources, with an uncertainty of about a factor of 2. In Figure~\ref{fig:luminosities} we show the observed and intrinsic 2--10~keV luminosities in the context of empirical correlations between $L_{\rm 2-10~keV}$ and $L_{\rm MIR}$ \citep{lutz+2004,fiore+2009,gandhi+2009}.

One notable difference between NGC~1320 and IC~2560 is that the latter is significantly less luminous in the hard X-ray bands, likely due to the much steeper intrinsic continuum suggested by our best-fit reflection model. Partly because of the obscuration so high that even the hard X-ray continuum is substantially suppressed and partly due to the steepness of the continuum, IC~2560 deviates significantly from the empirically determined distribution of observed luminosity ratios for Seyfert~2 nuclei with hard X-ray fluxes from \swiftbat$\!\!$. For example, \citet{lamassa+2010} find\footnote{No standard deviation was given, so we roughly estimate it based on standard deviations of other similar samples.} $\log(L_{\rm 14-195~keV}/L_{\rm MIR})=-0.8\pm0.3$ and $\log(L_{\rm 14-195~keV}/L_{\rm [O\,III]})=2.7\pm0.6$, compared to $-1.6$ and $1.3$ for IC~2560, respectively. NGC~424 and NGC~1320 deviate from those distributions by no more than 1.5 standard deviations, except for the case of $L_{\rm 14-195~keV}/L_{\rm [O\,III]}$ for NGC~424. Again, this may be due to the photon index being steeper than the population mean, however, it may also be related to the unusual optical spectrum of NGC~424, which shows features of both Type~1 and Type~2 AGN (see Table~\ref{tab:basic_data} and \S\,\ref{sec:discussion-geometry}).

NGC~424 is the most luminous of the three AGN in our sample. One of the spectral models that fully accounts for the features in its joint \xmmnewton and \nustar spectrum is a two-component \mytorus model consisting of an absorbed power law in addition to the reflected component. However, simple consideration of energetics essentially rules out the possibility that the absorbed power law represents the intrinsic continuum observed directly, absorbed along the line of sight to the nucleus. In the broad 14--195~keV band the de-absorbed intrinsic continuum luminosity is less than half of the reflected component luminosity ($1.3\times10^{42}$~erg~s$^{-1}$ compared to $2.9\times10^{42}$~erg~s$^{-1}$). With the high-energy cut-off just above this band, at 200~keV, such a continuum simply cannot provide sufficient photon flux to produce the observed spectrum. Furthermore, the infrared data, the internal normalization of the \mytorus model components, and a comparison to similar obscured AGN from the literature (e.g.,\,\citealt{matt+1997,matt+2004,burlon+2011,yaqoob-2012,arevalo+2014}) suggest that the intrinsic continuum should be nearly two orders of magnitude stronger in order to produce the observed reflected emission. This absorbed power law component could be well explained in a scenario in which the absorbed continuum comes from optically thin Thomson scattering of the nuclear continuum by free electrons, seen through an off-nuclear ``window'' of lower column density than that of the predominantly Compton-thick material. Such a scattered component is expected to contain $\sim$1--10\% of the power of the intrinsic continuum (e.g.,~\citealt{moran+2000}), which agrees well with our model.

The hypothesis that the intrinsic power law continuum is significantly stronger than what the absorbed power law component of \mytorus would imply is fully consistent with multi-wavelength properties of NGC~424 and its mid-infrared and [O\,III] luminosities in particular. In comparison with the published distributions for Seyfert~2 AGN samples, $\log(L_{\rm 2-10~keV}/L_{\rm MIR})=-2.4$ and $\log(L_{\rm 2-10~keV}/L_{\rm [O\,III]})=0.2$ make NGC~424 a severe outlier and suggest that it would take a boost of nearly two orders of magnitude from the observed luminosity to make its properties typical for a Seyfert~2 nucleus. We therefore estimate that the intrinsic 2--10~keV luminosity of NGC~424 is $\gtrsim2\times10^{43}$~erg~s$^{-1}$. Based only on data below 10~keV, \citet{marinucci+2011} infer a somewhat lower luminosity of $4\times10^{42}$~erg~s$^{-1}$. Using broad-band data from \asca and \bepposax (0.6--100~keV) \citet{iwasawa+2001} argued that under reasonable assumptions on geometry the intrinsic 2--10~keV luminosity could be as high as $2\times10^{43}$~erg~s$^{-1}$. With that luminosity NGC~424 would closely match the intrinsic relation between $L_{\rm 2-10~keV}$ and $L_{\rm MIR}$ from \citet{lutz+2004}, \citet{fiore+2009} and \citet{gandhi+2009} (see Figure~\ref{fig:luminosities}).

The adopted intrinsic luminosities may be converted to bolometric ones using empirical calibrations (e.g.,\,\citealt{marconi+2004,lusso+2012}). We estimate the bolometric correction factors for both the mid-IR and 2--10~keV bands to be 10--20, not including the considerable intrinsic uncertainty of a factor $\simeq$3. The bolometric luminosity is then $\sim7\times10^{44}$~erg~s$^{-1}$ for NGC~424 and $\sim2\times10^{44}$~erg~s$^{-1}$ for NGC~1320 and IC~2560. Combined with the measured masses of the SMBH harbored by the AGN we can estimate their Eddington fraction, $L_{\rm bol}/L_{\rm Edd}$, where $L_{\rm bol}$ is the bolometric luminosity and $L_{\rm Edd}$ is the Eddington luminosity. The mass of the SMBH in IC~2560 has been measured dynamically using the water megamasers \citep{ishihara+2001}, while for NGC~424 and NGC~1320 it was infered from measurements of stellar velocity dispersion via the $M_{\rm \tiny SMBH}-\sigma_{*}$ relation \citep{bian+gu-2007}. Using those values we infer Eddington fractions of $\sim$5\% for NGC~424 ($M_{\rm \tiny SMBH}=6.0\times10^7$~\msun) and NGC~1320 ($M_{\rm \tiny SMBH}=1.5\times10^7$~\msun), and $\sim$30\% for IC~2560 ($M_{\rm \tiny SMBH}=2.9\times10^6$~\msun). The uncertainties on these values are likely up to a factor of 5, but it is encouraging that the quantities are consistent with empirical correlations observed in large AGN samples (e.g.,\,\citealt{vasudevan+fabian-2009,lusso+2012}). We also find that the AGN with the highest Eddington fraction, IC~2560, has the steepest photon index, in accordance with the statistical relation found by \citet{brightman+2013} and consistent with the spread observed for individual sources in their sample.

\subsection{Constraints on the Geometry\\ of the Obscuring Material} 

\label{sec:discussion-geometry}

By modeling the hard X-ray spectra of NGC~424, NGC~1320 and IC~2560 we have unambigously confirmed that the dominant component in their hard X-ray spectra is a reflection of the nuclear continuum from distant, cold material. Although minor contributions from heavily absorbed intrinsic continua (at the $\lesssim$10\% level) cannot be completely ruled out with the data from short \nustar exposures presented here, it is clear that the sources are reflection-dominated and therefore that their nuclear regions must be heavily obscured by Compton-thick material. The type of reflection spectrum that fits all of the data best is the ``face-on'' component of the \mytorus model, which can be envisioned as the component that remains when the observer looks down the axis of symmetry of the torus, but with the central region, from where the intrinsic continuum is emitted, blocked by a completely opaque patch. According to \citet{yaqoob-2012}, this component closely approximates the spectrum one would observe in the case of a uniform Compton-thick torus being tilted towards the observer just enough for the far side of the torus to be visible in reflected light, but not enough to allow the nuclear continuum to be observable over the closer edge of the torus (e.g.\,\citealt{matt+2003b}).

Although the inclination limits for the situation described above depend on the scale height of the gas and dust distribution, one can argue that if the obscuring material really does reside in a toroidal structure, its orientation with respect to the observer would need to be close to edge-on. This orientation is indeed plausible given that water megamaser emission has been observed in two out of three targets \citep{braatz+1996,ishihara+2001,kondratko-2007,zhang+2010}, with IC~2560 being well established as a classical disk megamaser (see, e.g.,~\citealt{tilak+2008} for a discussion of the possible disk/torus and masing clouds' geometry). It is worth noting that the type of reflection spectrum that fits the \nustar data (both ``face-on'' \mytorus and \pexrav models) is more characteristic of surface scattering than scattering through dense material, which argues towards a tilted edge-on and possibly clumpy torus scenario, rather than complete enclosure in a spherical distribution of material. The \mytorus model used here is uniform, has sharp edges and a fixed covering fraction of $1/2$. These are approximations made in order to simplfy the model calculation -- astrophysical tori are likely neither uniform, nor have an outer solid surface. For those reasons, a blind application of the literal torus in which one would fit for all free parameters of the model (especially inclination) may not be reasonable or even recommended \citep{yaqoob-2012}.

From the multi-wavelength data presented in \S\,\ref{sec:discussion-luminosities} we infer a reflection efficiency in the 2--10~keV band of the order of 1\%, broadly consistent with any torus orientation \citep{ghisellini+1994}. In order to achieve the observed reflection dominance, the material surrounding the source must intercept a relatively large fraction of the nuclear luminosity (i.e.,~large covering fraction), while the side of the torus closer to the observer covers the direct line of sight with only a small global covering fraction -- letting the bulk of the backside reflection spectrum pass through unimpeded. A compelling physical scenario may be that of a clumpy torus \citep{nenkova+2008,elitzur-2012}. In that case the central source would be surrounded by a large number of individual high-column-density clouds with a large global covering factor, but also plenty of essentially clear lines of sight for their reflection spectra to reach the observer. In the clumpy torus picture the complete extinction of the intrinsic continuum along the line of sight to the nucleus could be explained by a large inclination angle and larger density of clouds in the equatorial plane of the system, which is again consistent with the observations of megamasers. 

Torus clumpiness also offers a natural explanation for the weak absorbed power law component required to model the NGC~424 spectrum: along some off-nuclear lines of sight there could be openings with relatively low column density ($\sim10^{23-24}$~cm$^{-2}$, compared to the surrounding $\gtrsim10^{25}$~cm$^{-2}$), through which we observe a weak continuum produced by optically thin Thomson scattering of the nuclear continuum on free electrons. Those free electrons are located in, or above, the opening of the torus and scatter the intrinsic continuum and the broad line region photons into our line of sight. This mechanism has already been invoked to explain the weak broad lines (e.g.,~H$\alpha$, H$\beta$, Fe~II) seen in the optical spectrum of NGC~424, which is otherwise typical of a Seyfert~2 AGN \citep{murayama+1998}, and the broad emission lines detected in polarized light \citep{moran+2000}, typical of Type~1 AGN with a ``hidden'' broad-line region. With newer data it is becoming possible to go beyond the simplistic Type~1 and Type~2 classification and more directly probe the geometry of the obscuring material. We have shown in previous sections that the infrared and the soft X-ray data, and especially the hard X-ray data from \nustar$\!\!$, unambigously confirm the heavy obscuration of the central source in NGC~424 regardless of the optical classification.

Using high angular resolution interferometric data, \citet{hoenig+2012} find that the dominant contribution to the nuclear mid-infrared flux comes from relatively cold optically-thin dust in a structure elongated in the direction of the torus axis of symmetry (as determined from the spectropolarimetric data of \citealt{moran+2000}). This observation suggests an interesting dust distribution in which optically thin clouds line the torus opening and possibly reach out to the narrow-line region (see Figure~9 in \citealt{hoenig+2012}). There is a possibility that a fraction of the optically thick clumps that predominantly reside in the torus get entrained in a wind together with the optically thin clouds that dominate the mid-infrared output; this is broadly consistent with our spectral modeling (i.e.~symmetrically broadened neutral, and Doppler-shifted ionized iron lines; see \S\,\ref{sec:modeling3-ngc424}).

The example of NGC~424 highlights the importance of modeling the hard X-ray spectra of Compton-thick AGN in synergy between data from different spectral bands in order to probe the unresolved AGN structure. The high degree of similarity with the Circinus Galaxy and NGC~1068 in both the interferometric mid-infrared data \citep{mason+2006,reunanen+2010,hoenig+2012} and the hard X-ray spectra from \nustar (\citealt{arevalo+2014}; Bauer \etal$\!\!$, {\it in prep.\ }) may be indicating that these objects are not merely isolated special cases. Due to the lack of similar data for NGC~1320 and IC~2560, at this point we can only speculate that they may exhibit a similar sort of geometry as NGC~1068, Circinus and NGC~424, which are brighter and have higher-quality broadband data.

Further evidence for clumpiness of the obscuring matter surrounding the nuclei in general can be found in observations of dramatically changing column density in some nearby AGN (e.g.,\,\citealt{matt+2003b,risaliti+2010,rivers+2011,walton+2014}), as well as the optical classification changes observed in at least a dozen AGN up to date (see, e.g.,\,\citealt{shappee+2013}). For NGC~424 and IC~2560 no significant variaibility in absorption has been observed in the literature, but an optically thin scattering medium has been invoked in modeling of their soft X-ray spectra in some observations \citep{matt+2003a,tilak+2008}. We observe no spectral changes between three epochs of NGC~1320 observations and the data are currently not of sufficient quality to constrain a possible Thomson-scattered continuum component. Hard X-ray variability could in principle be used to further constrain the physical size of the torus (e.g.,~\citealt{mattson+weaver-2004}), which might be possible with future repeated observations. The moderately large sample of nearby heavily obscured AGN currently surveyed with \nustar will provide the high-quality hard X-ray data needed for such work in the near future.

\begin{figure} 
\begin{center}
\includegraphics[width=0.98\columnwidth]{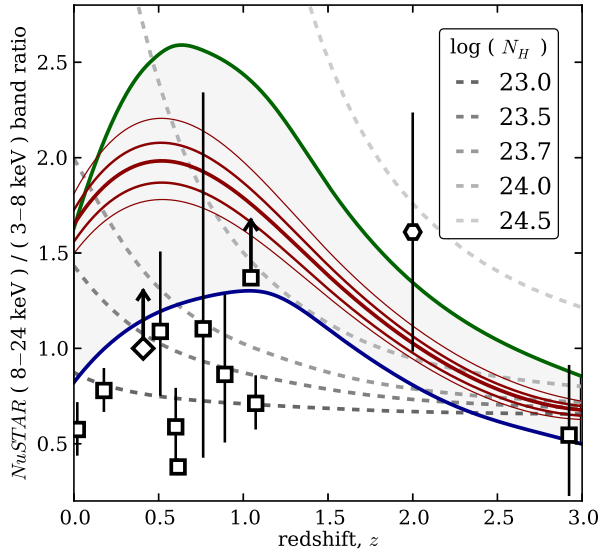}
\caption{ Observed count rate ratio for the \nustar 3--8 and 8--24~keV bands for some simple models at different redshift. The dashed lines in the background show tracks for a simple absorbed power law model (assuming $\Gamma=1.8$) for a range of column densities along the line of sight to the nucleus. Thick colored lines correspond to tracks of best-fit models for NGC~424 (red), NGC~1320 (green) and IC~2560 (blue). The differences between the three thick colored tracks are mostly due to different intrinsic photon indices; from 1.9 for NGC~1320 to 2.55 for IC~2560 (2.3 for NGC~424). With thinner red lines we show the effect of changing the intrinsic photon index of NGC~424 by 0.05 and 0.1 in either direction. Overplotted are data for the first 10 detected \nustar serendipitous sources (squares, from \citealt{alexander+2013}) and a lower limit for a heavily obscured SDSS quasar (diamond) presented in \citep{lansbury+2014}. Any source with a low count rate and a band ratio within the shaded grey area (or higher) may be considered a Compton-thick AGN candidate. However, the obscuration would need to be inferred from the mid-infrared, or other multi-wavelength data. \label{fig:redshifting}}
\end{center}
\end{figure} 

\subsection{Broader Context: Identification and\\ Properties of Obscured AGN} 

\label{sec:discussion-context}

The three sources presented here join a relatively small sample of nearby Compton-thick AGN that have been studied spectroscopically in some detail over the broad 0.2--200~keV energy range (e.g.,\,\citealt{maiolino+1998,matt+2000}). \nustar opens up an opportunity to extend our empirical knowledge towards less luminous and more distant Compton-thick AGN, and ultimately understand their properties better. Early results from the \nustar extragalactic surveys (\citealt{alexander+2013}; Mullaney \etal$\!\!$, {\it in prep.}; Civano \etal$\!\!$, {\it in prep.}) already include candidate conterparts to these local AGN at higher redshift (e.g., $z\approx0.5$, \citealt{lansbury+2014}; {$z\approx2$}, \citealt{delmoro+2014}, \citealt{stern+2014}). Short $\sim$20-ks \nustar observations of obscured luminous quasars at low redshift ($L_{\rm bol}\gtrsim10^{45}$~erg~s$^{-1}$, $z\lesssim0.2$; \citealt{gandhi+2014}) may reach sufficient photon statistics to enable modeling similar to that presented in this paper. However, for most of the high-redshift sources one is limited to using band ratios as indicators of their spectral shape. In Figure~\ref{fig:redshifting} we show the ratio of counts in the \nustar 3--8 and 8--24~keV bands for $z>0$ Compton-thick reflection-dominated AGN ($N_{\rm H,A}>10^{25}$~cm$^{-2}$) with X-ray spectra equal to those of the three AGN presented in this paper. Their tracks cross those of Compton-thin AGN (obscured by $N_{\rm H,A}\sim10^{23-24}$~cm$^{-2}$), showing that information beyond the band ratio is needed in order to isolate Compton-thick AGN. The distinction between Compton-thick and Compton-thin obscuration is of key importance in population studies (e.g., the distribution of $N_{\rm H,A}$ in obscured AGN) and understanding of the CXB, since the AGN obscured by $N_{\rm H,A}\lesssim10^{24}$~cm$^{-2}$ contribute significantly more hard X-ray radiation per source \citep{comastri-2004,gilli+2007}.

A three-band X-ray approach, similar to the one recently proposed by \citet{iwasawa+2012}, or \citet{brightman+nandra-2013}, would provide better means of discrimination between the levels of obscuration. However, at low number counts the uncertainties on band ratios in two bands can already be large, and dividing the data into three bands may not be beneficial. The possible confusion outlined in Figure~\ref{fig:redshifting} is aggravated by the fact that a spread in intrinsic photon indices exists in the population, as demonstrated by the differing photon indices of the three AGN presented here. Compared to the observed distribution of $\Gamma$ with a mean of 1.9 and an intrinsic width of a 0.3 (e.g.\,\citealt{brightman+nandra-2011a,burlon+2011,ballantyne-2014}), the photon indices determined from spectral analysis of the \nustar data span almost the full observed range, but do not represent extreme sources. The constant difference of approximately 0.4 between photon indices determined using the \pexrav models and those determined using \mytorus can be understood as arising from the different geometries and parameters assumed; we explore those differences in detail in a forthcoming publication (Brightman \etal$\!\!$, {\it in prep.}). As obscured AGN studies expand to higher redshifts and lower luminosities, the variance of different systems will only increase, which further motivates establishing good spectral templates based on nearby objects such as the ones presented here.

The typically low signal-to-noise ratio of sources identified in the blank-field \nustar surveys makes it difficult to distinguish different AGN types based only on band ratios and it is therefore crucial to consider multi-wavelength data. For example, any source found to occupy the grey area in Figure~\ref{fig:redshifting} has a spectral slope consistent with a heavily Compton-thick AGN, but only those with an extreme $L_{\rm 2-10~keV}/L_{\rm MIR}$ (as discussed in \S\,\ref{sec:discussion-luminosities}) should be considered true candidates. A similar concept of Compton-thick AGN selection was employed by \citet{severgnini+2012}. Many other studies of empirical multi-wavelength correlations and indirect obscuration indicators have been conducted in the past (e.g.,~\citealt{mulchaey+1994,lutz+2004,alexander+2008,fiore+2009,lamassa+2010,zhang+2010,georgantopoulos+2011,goulding+2011,iwasawa+2012}), directly or indirectly leading towards better selection criteria. With our small sample of three, we can identify the megamaser-biased selection of Compton-thick AGN in \citet{zhang+2010} as potentially inadequate, as none of our sources (despite two of them hosting water megamasers), would be classified as Compton-thick based on their multi-wavelength selection criteria. Perhaps owing to the similarity and simplicity of their spectra, all three sources generally pass the selection criteria proposed in the literature.

In this paper we use the empirical correlations, mainly between $L_{\rm 2-10~keV}$ and $L_{\rm MIR}$, in two ways: (i) assuming they hold for our sources, we infer that the intrinsic 2--10~keV luminosity has been suppressed by a factor of $\sim$100, and (ii) having determined that the X-ray spectra are dominated by reflection, which is typically assumed to be of the order of 1\% efficient in the 2--10~keV band, we verify that our sources still match the intrinsic relations and the Compton-thick AGN selection criteria. In that process we construct a complete self-consistent picture for each of our targets in order to provide a valuable benchmark for future multi-wavelength studies. With more precise measurements, better selection, tighter correlations and larger samples it will become possible to constrain the geometry of the circumnuclear material on statistical grounds and provide new tests for the contending AGN models (e.g.,\,\citealt{ghisellini+1994,urry+padovani-1995,elitzur-2012}).

\section{Summary \& Conclusions} 

\label{sec:conclusion}

In this paper we present hard X-ray spectroscopy of three highly obscured, nearby Seyfert nuclei: NGC~424, NGC~1320, and IC~2560. \nustar observations unambiguosly confirm that they are among the most obscured AGN in the local universe: the obscuration of the nuclei is well into the Compton-thick regime, where the nuclear continuum is suppressed to the extent that the distant/cold reflection component dominates the X-ray spectrum above 3~keV. Although hard X-ray data from nearly obscuration-unbiased surveys with \swiftbat and \integral were previously able to provide basic fluxes and spectral shapes, with \nustar it is possible study moderately large samples of hard X-ray selected AGN spectroscopically. With better understanding of local heavily obscured examples and their multi-wavelength properties we aim to achieve greater reliability in identifying their more distant counterparts. This will ultimately lead to improved indirect indicators that are often the only tools available for high-redshift AGN studies.

From the analysis of the quasi-simultaneous \nustar and \swiftxrt observations, as well as archival \xmmnewton and multi-wavelength data, we find the following:
\begin{itemize}
  \item All three sources show strong fluorescent iron lines and prominent Compton humps, as expected for reflection spectra. They stand out from the \nustar sample of local hard X-ray selected AGN by their hard effective photon indices and very high reflection strength, if modeled with simple spectral models. From a preliminary analysis of the ongoing \nustar survey of nearby AGN, we estimate that similar sources constitute approximately 10\% of that currently incomplete sample.
  \item Detailed modeling reveals that the X-ray spectra above 3~keV are dominated by reflection components and that no contributions from heavily absorbed intrinsic continua are formally required by the data. We thus infer that in all three cases the intrinsic continuum is obscured by $N_{\rm H,A}>5\times10^{24}$~cm$^{-2}$ and find $N_{\rm H,R}\gtrsim3\times10^{24}$~cm$^{-2}$ for the column density of the reflecting material. The dominance of the reflection component is further supported by the strong fluorescent Fe~K$\alpha$ lines with equivalent width of 1--2.5~keV.
  \item As the intrinsic X-ray continua are not observed in transmission, we estimate their luminosities from multi-wavelength data, and infer a reflection efficiency of the order of 1\% in the 2--10~keV band. Based on that efficiency and the surface-type reflection that fits the data best we argue for an edge-on clumpy torus geometry. Further studies of statistically representative samples of Compton-thick AGN are needed to constrain the typical reflection efficiency, which is currently only assumed to be of the order of 1\% in CXB models.
  \item Considering the extension of Compton-thick AGN studies to the high-redshift, low-count regime, we show that the band count ratio in the \nustar bandpass is not a good discriminator between mildly and heavily Compton-thick sources at $z>0$ and advocate usage of mid-infrared data to infer obscuration or reflection dominance. In that context, we use archival multi-wavelength data to verfy that (i) our sources obey intrinsic luminosity relations derived empiricaly from large AGN samples, and (ii) they would not miss being classified as Compton-thick AGN using most of the selection techniques proposed in the literature.
\end{itemize}

\acknowledgements
The authors thank the anonymous referee for useful comments which have improved the manuscript. M.\,B. acknowledges support from the International Fulbright Science and Technology Award. A.\,C. acknowledges support from ASI-INAF grant I/037/012/0-011/13. M.\,K. gratefully acknowledges support from Swiss National Science Foundation Grant PP00P2\_138979/1. This work was supported under NASA Contract No.~NNG08FD60C, and made use of data from the \nustar mission, a project led by the California Institute of Technology, managed by the Jet Propulsion Laboratory, and funded by the National Aeronautics and Space Administration. We thank the \nustar Operations, Software and Calibration teams for support with the execution and analysis of these observations. This research has made use of the \nustar Data Analysis Software (NuSTARDAS) jointly developed by the ASI Science Data Center (ASDC, Italy) and the California Institute of Technology (USA). This research made use of the XRT Data Analysis Software (XRTDAS), archival data, software and on-line services provided by the ASDC. This research has made use of NASA's Astrophysics Data System.

{\it Facilities:} \nustar$\!\!$, \swift$\!\!$, \xmmnewton

{} 

\end{document}